\documentclass[11pt]{article}

\usepackage[margin=1.0in]{geometry}
\usepackage{authblk}
\usepackage{background}

\usepackage{amsmath}
\usepackage{amssymb}
\usepackage{bm}
\usepackage{color,soul}
\usepackage{hyperref}
\usepackage{listings}
\usepackage{siunitx}
\usepackage{subcaption}

\usepackage{algorithm}
\usepackage[noend]{algpseudocode}
\algrenewcommand\algorithmicindent{1.0em}

\definecolor{lightlightgray}{gray}{0.98}
\definecolor{lightslategray}{RGB}{119,136,153}
\lstdefinestyle{myC++}{
	language=C++,
	basicstyle=\ttfamily,
    backgroundcolor=\color{lightlightgray},
    frame=tb, 
    showstringspaces=false, 
    numbers=left, 
    tabsize=2,
    breaklines=true,
    commentstyle=\color{lightslategray},
    keywordstyle=\color{red},
    keywordstyle=[2]\color{green},
    keywordstyle=[3]\color{blue},
    morekeywords=[1]{size_t},
    keywords=[2]{__global__, __syncthreads, blockDim, blockIdx, 
    threadIdx, gridDim, __device__, __forceinline__},
    keywords=[3]{parallel_for, RangePolicy, atomic_fetch_add, Serial, OpenMP, Cuda, LayoutLeft, LayoutRight, View}
}
\SetBgContents{Preprint}
\SetBgScale{1}
\SetBgAngle{0}
\SetBgOpacity{1}
\SetBgColor{black}
\SetBgPosition{current page.south west}
\SetBgHshift{1.0in}
\SetBgVshift{0.62in}

\begin{document}
\title{Performance portable ice-sheet modeling with MALI}

\author[1,*]{Jerry Watkins}
\author[1]{Max Carlson}
\author[2,**]{Kyle Shan}
\author[1]{Irina Tezaur}
\author[3]{Mauro Perego}
\author[3]{Luca Bertagna}
\author[4,**]{Carolyn Kao}
\author[5]{Matthew J. Hoffman}
\author[5]{Stephen F. Price}

\affil[1]{\footnotesize Sandia National Laboratories,  Quantitative Modeling \& Analysis Department, Livermore, CA, USA.}
\affil[2]{\footnotesize Micron Technology, Boise, ID, USA.}
\affil[3]{\footnotesize Sandia National Laboratories, Center for Computing Research, Albuquerque, NM, USA.}
\affil[4]{\footnotesize TSMC, Hsinchu, Taiwan.}
\affil[5]{\footnotesize Los Alamos National Laboratory, Fluid Dynamics and Solid Mechanics Group, Los Alamos, NM, USA.}
\affil[*]{\footnotesize Corresponding author; email: {\tt jwatkin@sandia.gov}.}
\affil[**]{\footnotesize All work completed as a student at Stanford University.}

\begin{titlepage}
\maketitle
\thispagestyle{empty}
\begin{abstract}
High resolution simulations of polar ice-sheets play a crucial role in the ongoing effort to develop more accurate and reliable Earth-system models for probabilistic sea-level projections. These simulations often require a massive amount of memory and computation from large supercomputing clusters to provide sufficient accuracy and resolution. The latest exascale machines poised to come online contain a diverse set of computing architectures. In an effort to avoid architecture specific programming and maintain productivity across platforms, the ice-sheet modeling code known as MALI uses high level abstractions to integrate Trilinos libraries and the Kokkos programming model for performance portable code across a variety of different architectures. In this paper, we analyze the performance portable features of MALI via a performance analysis on current CPU-based and GPU-based supercomputers. The analysis highlights performance portable improvements made in finite element assembly and multigrid preconditioning within MALI with speedups between 1.26--1.82x across CPU and GPU architectures but also identifies the need to further improve performance in software coupling and preconditioning on GPUs. We also perform a weak scalability study and show that simulations on GPU-based machines perform 1.24--1.92x faster when utilizing the GPUs. The best performance is found in finite element assembly which achieved a speedup of up to 8.65x and a weak scaling efficiency of 82.9\% with GPUs. We additionally describe an automated performance testing framework developed for this code base using a changepoint detection method. The framework is used to make actionable decisions about performance within MALI. We provide several concrete examples of scenarios in which the framework has identified performance regressions, improvements, and algorithm differences over the course of two years of development.
\end{abstract}
\end{titlepage}

\section{Introduction} \label{sec:intro}
\subsection{Motivation} \label{subsec:intro-motiv}
The Greenland and Antarctic ice-sheets contain the largest reserves of fresh water on Earth and have the greatest potential to cause changes in future sea level. The Special Report on the Ocean and Cryosphere in a Changing Climate (SROCC) \cite{oppenheimer2019sea} pointed to ice-sheets as one of the dominant contributors to rising and accelerating global mean sea level and stated that sea level will continue to rise for centuries due to mass loss from ice-sheets. Projections for future sea-level rise are dependent on the ability of ice-sheet models (ISMs) to simulate ice sheet mass loss, via a wide range of processes and instabilities, but major uncertainties in ice-sheet dynamics currently exist \cite{oppenheimer2019sea, Pattyn2017}.

In their 2007 Fourth Assessment Report, the Intergovernmental Panel on Climate Change (IPCC) defined clear deficiencies with ISMs ability to accurately capture dynamic processes and generally did not include these models when estimating future sea level rise \cite{randall2007climate}. Ice-sheet modeling improved dramatically with progress from many community supported ice-sheet models \cite{cornford2013adaptive,gagliardini2013capabilities,larour2012continental,rutt2009glimmer,winkelmann2011potsdam} and the IPCC's Fifth Assessment Report noted increased use of ISMs in climate models; however, major uncertainties remained \cite{flato2014evaluation}. Since then, there have been many studies which use multiple computational models to narrow the uncertainty but increasing computational cost remains a limiting factor \cite{edwards2021projected, Seroussi2020, Levermann2020, Goelzer2020b, Payne2021}. The SROCC noted that high resolution simulations without model simplifications are ultimately needed to obtain accurate projections of future global mean sea level \cite{oppenheimer2019sea}.

The increasing fidelity and resolution of ISMs pose significant computational challenges and demand their adoption of modern software techniques. This paper focuses on performance portable methods and optimizations that can be used to improve and maintain scalable performance in the presence of changing models, software and hardware.

\subsection{Performance portability} \label{subsec:intro-perf-port}
High resolution simulations of ice-sheet dynamics require a massive amount of memory and computation from large High Performance Computing (HPC) clusters, which coincidentally have undergone a dramatic change over the past decade. The current list of fastest supercomputers \cite{june2021} shows a diverse set of computing architectures, which typically include processors and accelerators from a variety of vendors. Software portability across these architectures is important for productivity, as the life cycle of a code base is typically much longer than the life cycle of individual supercomputers.

Heterogeneous compute nodes are also prevalent in new systems and will continue to be a challenge for software developers as the HPC community moves towards exascale \cite{heroux2020ecp}. The current strategy for HPC compute nodes utilizes two architectures, CPUs and GPUs, with fundamentally different design philosophies, to achieve high performance. CPUs are designed to minimize latency and are best utilized for sequential code performance while GPUs are designed to maximize memory and computational throughput in the presence of a sufficiently large and parallelizable work load. The latest CPUs also include vector processing units which can further improve computational throughput. In scientific computing, there are often many design choices that must be considered while developing software, and heterogeneous computing adds another design variable when choosing the best algorithms. Though it may be tempting to construct a highly optimized implementation for current HPC systems, this type of software development will become increasingly harder to maintain as future models, software and hardware become increasingly more complex. This motivates the need for fundamental abstractions to be present at the application level during code development. In response to these ongoing challenges in HPC, performance portability has grown to become crucial for simulating physical phenomena at high resolutions.

Even with the urgency of the challenge, there is still no consensus on a clear definition for the term ``performance portability'' \cite{neely2016doe,pennycook2021navigating}. In general, performance portability for an application means that a reasonable level of performance is achieved across a wide variety of computing architectures with the same source code. Here, ``performance" and ``variety" are also admittedly subjective. In \cite{pennycook2016metric,pennycook2017implications}, performance portability is quantified through efficiencies based on both application and architecture performance for a given set of platforms. In \cite{yang2018empirical}, this is extended to include the ``Roofline'' model which captures a more realistic set of empirically determined performance bounds. Since this paper focuses more on application level improvements, herein performance portability will be characterized by application execution time and scalability efficiencies for multicore/manycore processors and GPUs.

There have been a number of approaches to performance portability for application developers including directives such as OpenMP and OpenACC, and frameworks such as Kokkos \cite{edwards2014kokkos,trott2022kokkos}, RAJA \cite{hornung2014raja} and OCCA \cite{medina2014occa}. Performance portability for finite element analysis has also been executed on a variety of different software packages including Hiflow3 \cite{anzt2010hiflow3} and Firedrake/FEniCS/PyOP2 \cite{rathgeber2017firedrake,markall2013finite,rathgeber2012pyop2}.

As ISM codes evolve to be more robust, accurate, performant and portable on the latest HPC systems, a heavier burden is placed on software developers to support and improve functionality. Maintaining developer productivity is crucial for delivering on scientific discovery. Unfortunately, productivity is difficult to quantify, with a wide range of possible metrics \cite{forsgren2021space}. For scientific software development, version control and automated testing has been challenging to integrate \cite{kanewala2014testing,peng2021unit}, but have shown to be effective methods for improving productivity.

Automated testing becomes even more important as performance portable libraries and frameworks improve and expand their capabilities. As discussed in \cite{pennycook2021navigating}, these libraries and frameworks strive to improve developer productivity by reducing programming complexity and the need for platform-specific tuning while maintaining and improving performance. Staying up-to-date with these libraries and frameworks becomes crucial for maintaining an active code base which utilizes the latest HPC machines; however, maintaining performance portability in the presence of active development can be a difficult task. Small changes within a compiler, library, architecture, or code can cause dramatic changes to performance and performance deficiencies can be difficult to identify retroactively. Automated performance testing offers a means to improve productivity by reducing the time it takes to identify performance regressions and improvements to performance portability.

\subsection{Previous related work} \label{subsec:intro-related}
Traditionally for HPC, ISM codes relied solely on Message Passing Interface (MPI) libraries to achieve performance on supercomputers. MPI focuses on distributed memory parallelism, where memory may need to be communicated across multiple compute nodes. In \cite{gagliardini2013capabilities}, a 60\% weak scalability efficiency is computed on a set of Greenland ice-sheet meshes for the full Stokes solver in Elmer/Ice using 168 to 1092 cores. The computational component of the hybrid ``SSA+SIA" model in PISM is found to scale well on up to 1024 cores in \cite{dickens2015performance} but the I/O component is found to scale poorly. In \cite{fischler2021scalability}, low-overhead performance instrumentation is developed for the ``Blatter-Pattyn" model in ISSM and good scaling is found on up to 3072 cores. The study finds that matrix assembly and I/O begins to scale poorly and highlights the importance of continuous performance monitoring.

A code with MPI-only is not able to take advantage of the computational throughput available on shared memory architectures including compute nodes with dedicated GPUs. GPUs can provide substantial performance improvements to existing ISMs if properly utilized. In \cite{braedstrup2014ice}, a CUDA implementation of the ``iSOSIA" approximation is used to show that higher-order ice flow models can be significantly accelerated with NVIDIA GPUs. FastICE is introduced in \cite{rass2020modelling} as a parallel, GPU-accelerated full Stokes solver developed in CUDA, which utilizes a matrix-free method with pseudo-transient continuation. This is extended to a portable framework written in Julia in \cite{rass2021assessing} and a parallel efficiency over 96\% on 2197 GPUs is achieved.

In this work, the velocity solver in MALI, which uses the ``Blatter-Pattyn" model formulation, is extended to be performance portable using a multigrid preconditioned, Newton-Krvylov method where extensive improvements have been made to matrix assembly performance. The performance and scalability of MALI and its velocity solver is analyzed on multiple architectures including Intel Knights Landing (KNL) and NVIDIA V100 GPUs. The testing framework in MALI is also extended to include performance monitoring with automated detection of performance regressions and improvements using a unique changepoint detection method.

MALI (MPAS-Albany Land Ice, \cite{hoffman2018mpas}) is an ice-sheet model built on top of two main libraries: the MPAS (Model for Prediction Across Scales) library \cite{ringler2013},  written in Fortran and used for developing variable-resolution Earth system model components, and Albany, a C++ finite element code for solving partial differential equations \cite{salinger2016albany}. The performance of MALI is dominated by the solution of the first-order approximation to the Stokes equations (hereafter simply first-order velocity or first-order, see Section~\ref{sec:ISM} below); hence, the performance portability efforts described in this work have been mainly targeting the C++ implementation of these equations in Albany. We note that MALI can model several additional physical processes including the ice temperature evolution, subglacial hydrology and iceberg calving \cite{hoffman2018mpas}.

Albany uses high level abstractions to integrate Trilinos libraries \cite{heroux2005overview} and the Kokkos programming model \cite{edwards2014kokkos,trott2022kokkos} for performance portable code across a variety of different architectures. Albany follows an ``MPI+X" programming model, where MPI is used for distributed memory parallelism and the Kokkos library is used for shared memory parallelism. Kokkos provides abstractions for parallel execution and data management of shared memory computations in order to obtain optimal data layouts and hardware features, reducing the complexity of the application code. The performance portable implementation in Albany is described in detail in \cite{demeshko2018toward} where the authors highlight finite element assembly performance for Aeras, the atmospheric dynamical core implemented in Albany.

Albany Land Ice (ALI) is first introduced in \cite{tezaur2015albany} under the name Albany/FELIX. In \cite{tezaur2015scalability}, the scalability of the multigrid preconditioned, velocity solver is analyzed on up to 1024 cores. An initial study on the performance portability of the finite element assembly showed deficiencies in distributed memory assembly on GPU architectures in \cite{watkins2020study} but performance and scalability was reasonable among different architectures. 

In this paper, we highlight recent improvements to finite element assembly that eliminate previous deficiencies. We also begin analyzing a new, performance portable velocity solver in ALI and expand our performance analysis to MALI.

\subsection{Main contributions} \label{subsec:intro-main-cont}
The MALI code was developed in response to the growing challenges in developing a more accurate and efficient ISM \cite{hoffman2018mpas}. In this paper, the performance portable features of MALI are introduced and analyzed on the two supercomputing clusters: NERSC Cori and OLCF Summit. A changepoint detection method is also introduced and tested for automated performance testing on next generation architectures. The main contributions of this work are summarized as follows:
\begin{itemize}
\item Insights into the development of a performance portable, finite element code base using high-level abstractions from Trilinos libraries and the Kokkos programming model.
\item A description of new, performance-enhancing features introduced in MALI and an analysis demonstrating the expected improvements on different HPC architectures, including Intel Knights Landing (KNL) and NVIDIA V100 GPUs.
\item A weak scalability study and a demonstration of speedup over CPU-only simulations.
\item Insights into the development of a changepoint detection method for automated performance testing and demonstrations of tracking performance regressions, improvements, and differences between algorithms.
\end{itemize}
The methods introduced focus on improving performance portable modeling in MALI, but are extensible to other applications targeting HPC.

The remainder of this paper is organized as follows. Section~\ref{sec:ISM} introduces the ice-sheet equations relevant to our analysis. Section~\ref{sec:impl} gives a detailed overview of how these equations are implemented, solved and verified in MALI. In Section~\ref{sec:impr}, the methods used to achieve, improve and maintain performance portability in MALI are introduced. Lastly, Section~\ref{sec:res} provides three numerical examples that demonstrate the expected performance of MALI on HPC systems and the utility of automated performance testing.

\section{The governing ice-sheet equations} \label{sec:ISM}
In this section, the main equations governing the ice-sheet dynamics are briefly discussed. The section begins with a description of the first-order velocity equations and is followed by a description of the mass continuity equations. More information can be found in \cite{hoffman2018mpas, tezaur2015albany}. 
In this work, we will always assume a ``topologically extruded'' ice-sheet geometry, meaning that the ice-sheet geometry can be obtained by vertically extruding the two-dimensional (2D) basal area, according to the local ice thickness. A consequence of this assumption is that the margin of the ice-sheet is always vertical, though the ice thickness is typically small at grounded glacier termini.

At the ice-sheet scale, ice behaves as a highly viscous, shear-thinning, incompressible fluid and can be modeled by nonlinear Stokes equations. In this paper, a first-order approximation \cite{dukowicz2010consistent, schoof2010thin} of the Stokes equations is considered, often referred to as the ``Blatter-Pattyn" model \cite{blatter1995velocity, pattyn2003new} or the ``first-order'' model. The model is quasi-static with static velocity (momentum balance) equations coupled to a dynamic thickness (mass) equation. In conservative form, the three-dimensional (3D), first-order velocity equations are written as,
\begin{align}
\begin{split} \label{eq:FOStokes}
-\nabla \cdot \left(2\mu_e \dot{\bm{\epsilon}}_1 \right) + \rho g \frac{\partial s}{\partial x} = 0, \\
-\nabla \cdot \left(2\mu_e \dot{\bm{\epsilon}}_2 \right) + \rho g \frac{\partial s}{\partial y} = 0,
\end{split}
\end{align}
where $x$, $y$ and $z$ are spatial coordinates bounded by the ice domain $\Omega$, $\rho$ is the ice density, $g$ is gravitational acceleration and $s \equiv s\left(x,y\right)$ is the upper surface of the domain. The strain rates in Equation~\eqref{eq:FOStokes} are defined as the vectors,
\begin{align}
\begin{split}
\dot{\bm{\epsilon}}_1 = \left[2\dot{\epsilon}_{xx}+\dot{\epsilon}_{yy}, \dot{\epsilon}_{xy}, \dot{\epsilon}_{xz} \right]^T, \\
\dot{\bm{\epsilon}}_2 = \left[\dot{\epsilon}_{xy}, \dot{\epsilon}_{xx}+2\dot{\epsilon}_{yy}, \dot{\epsilon}_{yz} \right]^T,
\end{split}
\end{align}
where the components of the approximate strain rate tensor can be written as,
\begin{align}
\dot{\epsilon}_{xx} = \frac{\partial u}{\partial x}, \qquad \dot{\epsilon}_{yy} = \frac{\partial v}{\partial y}, \qquad \dot{\epsilon}_{xy} = \frac{1}{2}\left(\frac{\partial u}{\partial y}+\frac{\partial v}{\partial x}\right), \qquad \dot{\epsilon}_{xz} = \frac{1}{2}\frac{\partial u}{\partial z}, \qquad \dot{\epsilon}_{yz} = \frac{1}{2}\frac{\partial v}{\partial z},
\end{align}
and $u$ and $v$ are the ice velocity components in the direction of $x$ and $y$, respectively.

The effective viscosity, $\mu_e$, in Equation~\eqref{eq:FOStokes} is derived from Glen's flow law \cite{cuffey2010physics,nye1957distribution} and is written as,
\begin{equation} \label{eq:effvisc}
\mu_e = \frac{1}{2}A^{-\frac{1}{n}}\dot{\epsilon}_{e}^{\frac{1}{n} - 1},
\end{equation}
where $n$ is Glen's power law exponent and $\dot{\epsilon}_{e}$ is the effective strain rate, given by
\begin{equation} \label{eq:effeps}
\dot{\epsilon}_{e}^2 = \dot{\epsilon}_{xx}^2 + \dot{\epsilon}_{yy}^2 + \dot{\epsilon}_{xx}\dot{\epsilon}_{yy} + \dot{\epsilon}_{xy}^2 + \dot{\epsilon}_{xz}^2 + \dot{\epsilon}_{yz}^2.
\end{equation}
The flow law rate factor in Equation~\eqref{eq:effvisc} is strongly temperature dependent and is determined through an Arrhenius relation,
\begin{equation}
A = A_0 \exp\left(-\frac{Q}{RT^*}\right),
\end{equation}
where $A_0$ is a constant of proportionality, $Q$ is the activation energy for ice creep, $T^*$ is the ice temperature corrected for the pressure melting point and $R$ is the universal gas constant.

The boundary conditions are best described by partitioning the surface of the 3D ice-sheet domain into upper, lower and lateral surfaces,
\begin{equation}
\Gamma = \Gamma_s \cup \Gamma_{\beta} \cup \Gamma_l,
\end{equation}
where $\Gamma_s$ is the upper surface, $\Gamma_{\beta}$ is the lower surface and $\Gamma_l$ is the lateral surface.
The boundary conditions can then be defined as:
\begin{enumerate}
\item a homogeneous boundary condition on the upper surface (atmosphere pressure is neglected),
\begin{equation} \label{eq1}
\dot{\bm{\epsilon}}_1 \cdot \bm{n} = 0, \qquad \dot{\bm{\epsilon}}_2 \cdot \bm{n} = 0, \qquad \text{on } \Gamma_s,
\end{equation}
where $\bm{n}$ is the outwards facing normal vector;
\item a Robin boundary condition on the lower surface, representing a linear sliding law at the bed,
\begin{equation}
2\mu_e\dot{\bm{\epsilon}}_1 \cdot \bm{n} + \beta u, \qquad 2\mu_e\dot{\bm{\epsilon}}_2 \cdot \bm{n} + \beta v, \qquad \text{on } \Gamma_{\beta},
\end{equation}
where the basal sliding coefficient $\beta \equiv \beta\left(x,y\right)$ is non-negative where the ice is grounded and zero where the ice is floating;
\item a dynamic Neumann boundary condition at the ice margin accounting for the back pressure from the ocean where the ice is submerged (note that by convention $z=0$ represents the sea level),
\begin{align} \label{eq2}
\begin{split}
&2\mu_e\dot{\bm{\epsilon}}_1 \cdot \bm{n} - \rho g \left(s - z\right)\bm{n} = \rho_w g \max\left(z, 0\right)\bm{n}, \\
&2\mu_e\dot{\bm{\epsilon}}_2 \cdot \bm{n} - \rho g \left(s - z\right)\bm{n} = \rho_w g \max\left(z, 0\right)\bm{n}, \qquad \text{on } \Gamma_l,
\end{split}
\end{align}
where $\rho_w$ is the density of water and $z$ is the elevation above sea level.
\end{enumerate}

The steady velocity equations described above are coupled to a dynamic equation for the conservation of mass.  Specifically, as the ice-sheet evolves in time, mass continuity is enforced through the following equation,
\begin{equation} \label{eq:thickness}
\frac{\partial H}{\partial t} + \nabla \cdot \left(H\bm{\bar{u}}\right) = \dot{a} + \dot{b},
\end{equation}
where $H$ is ice thickness, $t$ is time, $\bm{\bar{u}}$ is depth-averaged velocity vector, $\dot{a}$ is surface mass balance and $\dot{b}$ is basal mass balance. The thickness equation is then used to evolve the geometry in time.
The ice temperature is held constant in time.

\section{Implementation in MALI} \label{sec:impl}
In this section we describe how the governing equations introduced in Section~\ref{sec:ISM} are discretized and implemented in MALI, focusing in particular on the C++ velocity solver model in ALI. We first give a high-level overview of the implementation and then provide a detailed description for the two more computationally expensive components relevant to the paper: the finite element assembly and the preconditioned linear solver. A brief description of the MALI testing framework follows.

\subsection{Overview} \label{subsec:impl-overview}
The ice thickness $H$ in Equation~\eqref{eq:thickness} is discretized in MPAS with a upwind finite volume method on an unstructured, two-dimensional, Voronoi grid \cite{hoffman2018mpas}. At every time step, the first-order velocity equations (Section~\ref{sec:ISM}) are solved in ALI \cite{tezaur2015albany}. Briefly, the equations are discretized using low-order nodal prismatic finite elements on a 3D mesh extruded from a triangulation dual to the MPAS Voronoi mesh. The discrete version of the velocity equations can be written in the compact form
\begin{equation} \label{eq:residual}
\mathcal F(U;\; \{\phi_i\}, \{\nabla \phi_i\}, H, \beta, \ldots) = 0.
\end{equation}
Here $U$ is the solution vector, containing the values of ice velocity at the mesh nodes. $\{\phi_i\}$ and $\{\nabla \phi_i\}$ are the sets of the basis functions and their gradients. $\mathcal F$ is a vector function of the solution $U$. $\mathcal F$ also depends on the basis functions and on fields like ice thickness $H$ and basal friction $\beta$. We refer to $\mathcal F(U;\; \cdot)$ as the \emph{residual}.

A damped Newton's method is used to solve the nonlinear discrete system \eqref{eq:residual}:
\begin{equation} \label{eq:linearsystem}
\mathcal J_{\mathcal F}^k\; \delta_{U}^{k+1}  = - \mathcal F(U^k), \qquad U^{k+1} = U^k + \alpha_k\, \delta_{U}^{k+1}.
\end{equation}
Here $\mathcal J_{\mathcal F}^k := \left . \dfrac{\partial \mathcal F}{\partial U} \right|_{U = U^k}$ is the \emph{Jacobian} matrix and $\alpha_k$ is the damping factor.
At each nonlinear iteration $k$, the linear system \eqref{eq:linearsystem} is solved with the GMRES method using the ``matrix-dependent semicoarsening-algebraic multigrid'' (MDSC-AMG)~\cite{tuminaro2016matrix} preconditioner.

\begin{figure}[htbp]
\includegraphics[width=\textwidth]{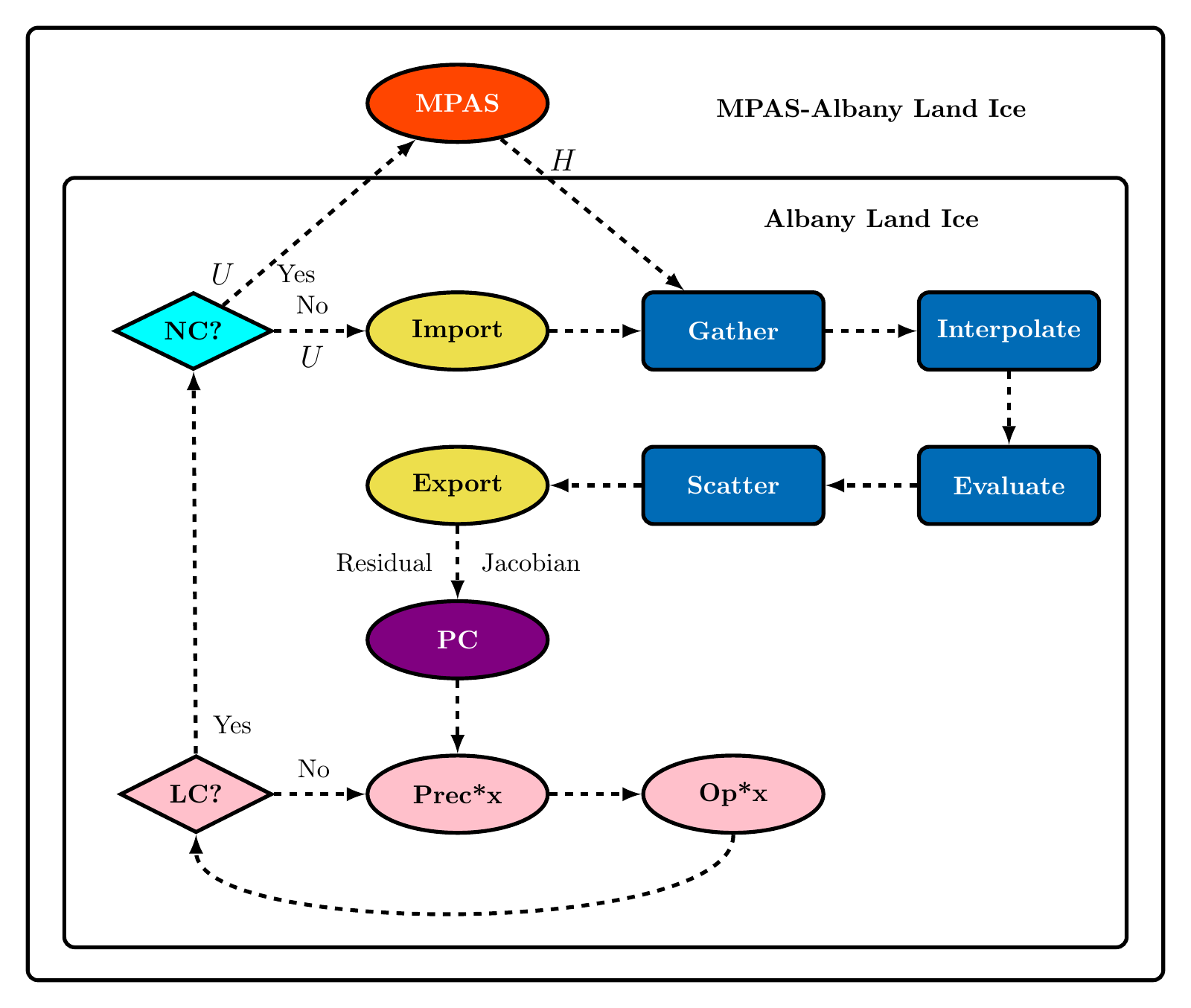}
\caption{A flow chart depicting the workflow in MALI, focusing on high-level components in the velocity solver in ALI. The shapes of the nodes are constructed to show performance relevant characteristics of the code while the colors are used to differentiate high-level abstractions. In this case, ellipses represent portions of code with MPI communication and rounded rectangles represent portions of code which do not require MPI communication. The conditionals, LC (Linear solver converged) and NC (nonlinear solver converged), are represented with diamonds. The red-orange node represents the computation required for explicit time stepping of ice thickness, $H$, in MPAS, the purple node represents preconditioner construction (PC), and the pink nodes represent the linear solver required to solve \eqref{eq:linearsystem}. Finite element assembly begins and ends with distributed memory assembly colored in yellow but also performs shared memory processes colored in blue.}
\label{fig:flow-chart}
\end{figure}

Figure~\ref{fig:flow-chart} shows a flow chart of ice-sheet dynamics in MALI, focusing on high-level components in the velocity solver in ALI.
Each node of the flow chart is described below with references to relevant Trilinos packages:
\begin{itemize}
\item {\bf {Import}}: Imports the ice velocity solution, $U$, from a nonoverlapping data structure where each MPI rank owns a unique part of the solution to an overlapping data structure where some data exists on multiple ranks. This gives each rank access to relevant solution data without any further communication. This is performed by Trilinos/Tpetra \cite{baker2012tpetra}.

\item {\bf {Gather}}: Gathers solution values from an overlapping data structure to an element local data structure where data is indexed according to element and local node. This process also includes the gathering of geometry and field data from MPAS. This is constructed using Trilinos/Phalanx \cite{pawlowski2012automatingI,pawlowski2012automatingII,salinger2016albany} and Trilinos/Kokkos \cite{edwards2014kokkos,trott2022kokkos}.

\item {\bf {Interpolate}}: Interpolates the solution and solution gradient from nodal points to quadrature points. Other field variables also require interpolation. This is constructed using Trilinos/Phalanx and Trilinos/Kokkos.

\item {\bf {Evaluate}}: Evaluates the residual, Jacobian and source terms of the first-order equations. These operators are templated in order to take advantage of automatic differentiation for analytical Jacobians using Trilinos/Sacado \cite{phipps2012efficient}. This also uses Trilinos/Phalanx and Trilinos/Kokkos.

\item {\bf {Scatter}}: Scatters residual and Jacobian values from an element local data structure to an overlapping data structure. This is constructed using Trilinos/Phalanx and Trilinos/Kokkos.

\item {\bf {Export}}: Exports the residual and Jacobian from an overlapping data structure to a nonoverlapping data structure where information is updated across MPI ranks. This global structure allows for efficient use of linear solvers and is performed by Trilinos/Tpetra.

\item {\bf {PC (Preconditioner Construction)}}: Constructs the MDSC-AMG preconditioner from the Jacobian matrix and is performed by Trilinos/MueLu \cite{berger2019muelu} and Trilinos/Ifpack2 \cite{prokopenko2016ifpack2}.

\item {\bf {Prec*x}}: Applies the preconditioner to the solution vector of the linear system and is performed by Trilinos/Belos \cite{bavier2012amesos2}, Trilinos/MueLu, Trilinos/Ifpack2.

\item {\bf {Op*x}}: Applies the Jacobian matrix to the solution vector of the linear system and is performed by Trilinos/Belos.

\item {\bf {LC? (Linear Solver Converged?)}}: The linear solver loop is converged when a fixed linear tolerance or a maximum number of iterations is reached. This is managed by Trilinos/Belos.

\item {\bf {NC? (Nonlinear Solver Converged?)}}: The nonlinear solver loop is converged when a fixed nonlinear tolerance is reached. This is managed by Trilinos/NOX \cite{nox-loca-website}.

\item {\bf {MPAS}}: Once the ice velocity, $U$, has fully converged, it is interpolated to MPAS cell edges and used to update $H$ using forward Euler. The new ice-sheet geometry is passed back into Albany Land Ice to re-compute $U$ for the next time step. This process is performed until a final time step is reached. A more detailed description of thickness evolution in MPAS can be found in \cite{hoffman2018mpas}.
\end{itemize}
By constructing high-level abstractions for solving nonlinear partial differential equations and utilizing Trilinos packages as components, application developers are able to apply existing performance portable algorithms which are actively supported and improved by experts. Developers are also able to utilize the same algorithms on multiple applications, allowing for greater impact and increased sustainability. 

\subsection{Finite element assembly} \label{subsec:impl-FEA}
\begin{figure}[htbp]
\centering
\includegraphics{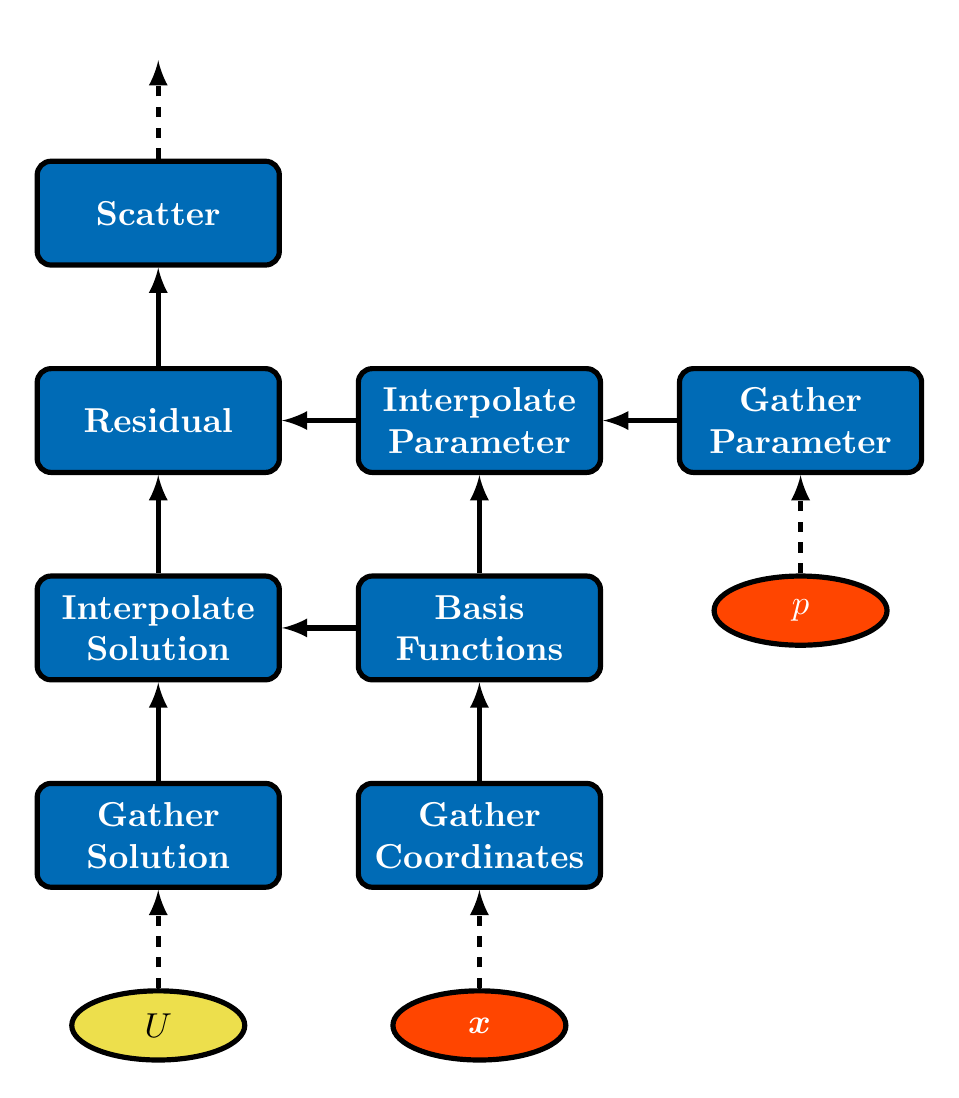}
\caption{Albany uses a directed acyclic graph (DAG) for finite element assembly. In this example, a global residual is constructed by interpolating the solution and a parameter onto quadrature points. Basis functions are also needed to complete the interpolation which depend on physical coordinates. In the case of the first-order equations, the solution is the ice velocity. One example of a parameter could be the surface height.}
\label{fig:phalanx-example}
\end{figure}

The finite element approach implemented in Albany is designed to easily incorporate multiple physics models with graph-based evaluation using the Trilinos/Phalanx package \cite{pawlowski2012automatingI,pawlowski2012automatingII,salinger2016albany}. The assembly is decomposed into a set of nodes called ``evaluators''. Evaluators have a specified set of inputs and outputs and are organized in a directed acyclic graph (DAG) based on dependencies. Figure~\ref{fig:phalanx-example} shows a simplified example of a DAG for the for finite element assembly in Albany.
The inherent advantage of using a DAG is the increased flexibility, extensibility and usability of using modular evaluators when performing finite element assembly. The DAG also provides potential for task parallelism. The disadvantage of using a DAG is that there is a potential for performance loss through code fragmentation and a static graph can also lead to repetition of unneeded data movement and computation. This is discussed in more detail in Section~\ref{subsec:impr-mem}.

Albany utilizes automatic differentiation to compute the Jacobian matrix and enable Newton-like methods for the solution of nonlinear partial differential equations. More in general, the finite element assembly in Albany is designed for embedded analysis using template-based generic programming \cite{pawlowski2012automatingI,pawlowski2012automatingII,salinger2016albany}. Embedded analysis, such as derivative-based optimization and sensitivity analysis, for partial differential equations requires construction of mathematical objects such as Jacobians, Hessian-vector products and derivatives with respect to parameters. Albany utilizes C++ templates and operator overloading to perform automatic differentiation using the Trilinos/Sacado package \cite{phipps2012efficient}.
Sacado provides multiple data types for storing the derivative components, each with their own relative advantages and disadvantages.
The {\bf \lstinline{DFad}} options sets the number of derivative components at runtime, and is hence the most flexible but least efficient option. The {\bf \lstinline{SLFad}} options sets the maximum number of derivative components at compile time, making this option both flexible and relatively efficient. The most efficient but least flexible option is {\bf \lstinline{SFad}}. For this option, the number of derivative components is set at compile time.
In MALI, we generally select the \lstinline{SFad} type whenever possible, so as to achieve the best possible performance. The difference in performance with the various options is most profound in a GPU run, due to the substantial cost of performing dynamic allocation on the GPU.

\begin{figure}[htbp]
\centering
\includegraphics[width=0.85\textwidth]{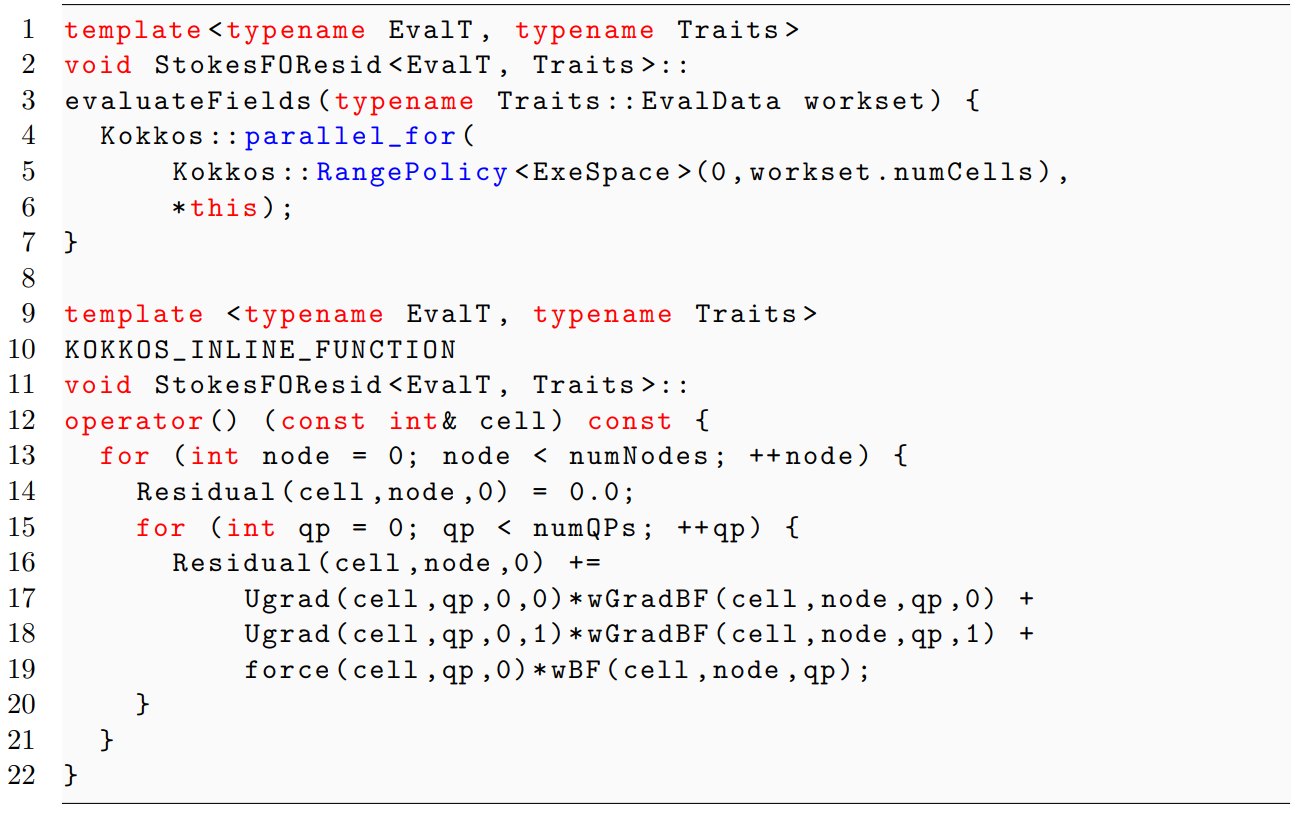}
\caption{The first-order local residual computation is performed in a Phalanx evaluator that uses Kokkos for shared memory parallelism and Sacado for automatic differentiation. This is a simplified example of the computation of a single term in the residual. When called with a \lstinline{double} \lstinline{EvalT} type, the routine returns the residual; when called with a \lstinline{Fad} \lstinline{EvalT} type, automatic differentiation is applied and a Jacobian is returned.}
\label{fig:kokkos-example}
\end{figure}

Finite element evaluators also contain Kokkos parallel execution kernels for performance portability on shared memory architectures \cite{demeshko2018toward,watkins2020study}. Kokkos utilizes memory and execution spaces to determine where memory is stored and where code is executed. Phalanx evaluators utilize \lstinline{MDField} with Kokkos \lstinline{View} for memory management and evaluators can be used as a Kokkos functor to perform parallel operations. Sacado operators have also been designed to work with Kokkos. Figure~\ref{fig:kokkos-example} shows a simplified example of an evaluator with Kokkos.

The Phalanx evaluation type is passed via the template parameter \lstinline{EvalT} and dictates whether a residual with a \lstinline{double} type or a Jacobian with a \lstinline{Fad} type is computed. A Kokkos \lstinline{RangePolicy} is used to parallelize over cells over an execution space, \lstinline{ExeSpace}. In this paper, only the \lstinline{Serial} and \lstinline{Cuda} executions spaces are used to differentiate between CPU and GPU execution but other execution spaces are also available. The properties of each case are described in more detail in \cite{demeshko2018toward,watkins2020study}

\subsection{Preconditioner for linear solve} \label{subsec:impl-linsol}
A primary challenge in simulating ice-sheets at scale is solving the linear system associated with a thin, high-aspect ratio mesh. It has been shown \cite{brown2013achieving, Issac:2015, tuminaro2016matrix} that multigrid methods can be used to address the challenges arising due to the anisotropic nature of the problem, although alternative methods have been proposed \cite{chen2019robust,heinlein2022frosch}. The performance of the linear solver is thus primarily dictated by the efficacy of the multigrid preconditioner.
The MDSC-AMG preconditioner introduced in \cite{tuminaro2016matrix} is specifically designed for ice-sheet meshes where a mesh is first constructed from 2D topological data and extruded in the vertical direction to construct a 3D mesh. The primary strategy for the multigrid hierarchy is to coarsen the fine mesh in the vertical direction until a single layer is reached and apply smoothed aggregation algebraic multigrid (SA-AMG) on the plane. Here, we are able to take advantage of the performance portable implementations of SA-AMG and point smoothers implemented in Trilinos/MueLu and Trilinos/Ifpack2.

\subsection{Testing} \label{subsec:impl-test}
Software quality tools are a central part of the Albany code base and are crucial for developer productivity \cite{salinger2016albany}. Rather than using a fixed release of Trilinos, ALI is designed to stay up-to-date with Trilinos' version of the day, to ensure that the code inherits the most up-to-date additions and improvements to Trilinos.
This requires a close collaboration between Albany and Trilinos developers and ensures rapid response to issues that might arise. The current nightly test harness includes unit, regression and performance tests on Intel and IBM multicore CPUs and NVIDIA GPUs and is monitored on a dedicated dashboard.

\section{Methods for improving and maintaining performance portability} \label{sec:impr}
In this section, we discuss the major enhancements made to MALI to both improve and maintain performance portability. We provide three examples of finite element assembly optimizations, which improved performance on both CPU and GPU systems. Memoization is utilized to avoid unnecessary data movement and computation from the MALI workflow. Optimizations in matrix assembly and boundary condition computation led to significant speedups on both CPU and GPU and a large reduction in memory usage. We also provide a brief description of a new, performance portable MDSC-AMG preconditioner implemented in Trilinos/MueLu and tuned for ice-sheet modeling. Lastly, we provide a description of an automated performance testing framework for identifying regressions, improvements and performance differences between algorithms.

\subsection{Memoization} \label{subsec:impr-mem}
A static DAG similar to the one shown in Figure~\ref{fig:phalanx-example} is executed when a new global residual or Jacobian is needed within the nonlinear solver. This leads to a repetition of unnecessary data movement and computation when input quantities do not change between calls of the DAG. A performance gain can be achieved by storing the results of expensive nodes in the DAG and returning the stored results when input quantities do not change. This process is known as memoization.

\begin{figure}[htbp]
\centering
\includegraphics{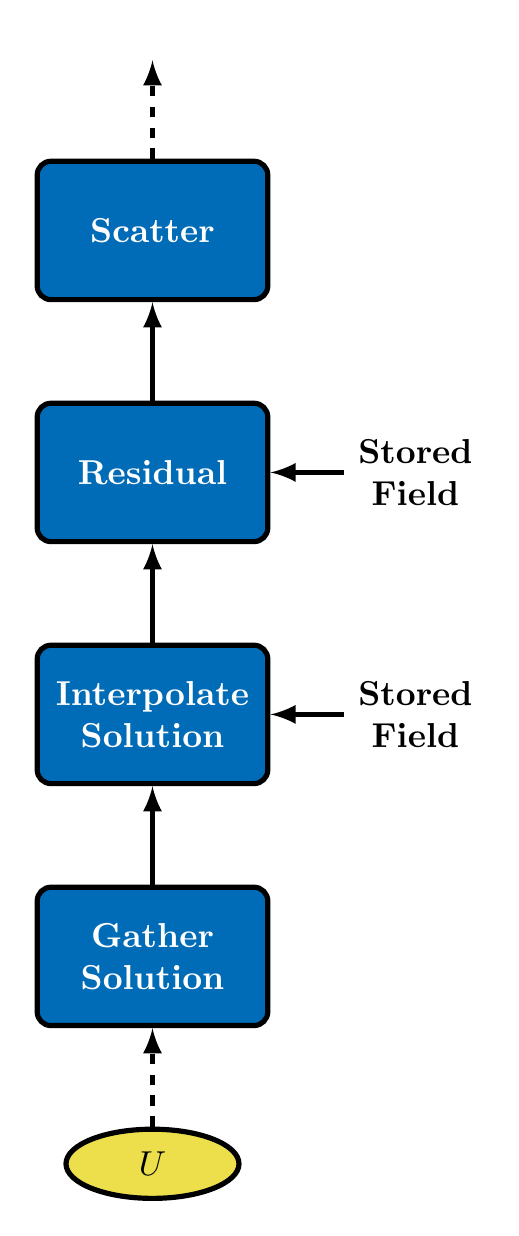}
\caption{Albany uses memoization to create a new DAG which only depends on changes to the solution. This avoids unnecessary data movement and computations when parameters and coordinates do not change. For examples, in the case of first-order equations, the surface height does not change at each nonlinear iteration.}
\label{fig:mem-example}
\end{figure}

In Albany, memoization is implemented by constructing a new DAG which only follows changes caused by the solution. Figure~\ref{fig:mem-example} shows an example of the new DAG created by performing memoization in Figure~\ref{fig:phalanx-example}.
The first call executes the original DAG while storing all intermediate quantities. Then, the new DAG is called by default while the original DAG is only called when there is a change to a parameter or the coordinates. An initial speedup of roughly 1.4 on CPUs and 1.5 on GPUs was found when analyzing finite element assembly performance relative to the assembly without memoization.

\subsection{Jacobian matrix} \label{subsec:impr-fecrs}
As discussed in Section \ref{sec:impl}, the finite element assembly of residual and Jacobian is performed on an ``overlapped'' distribution of the DOFs, while linear systems require a ``unique'' distribution of DOFs.

Until recently, Albany was using two separate Tpetra \texttt{CrsMatrix} objects for the Jacobian: an overlapped version for finite element assembly, and a non-overlapped version for linear solvers. An Export operation (involving MPI communication) was used to copy data between the overlapped and the unique matrices, migrating off-processor rows to their owner.

We improved this portion of the library by switching to the new Tpetra \texttt{FeCrsMatrix} objects, which can store overlapped and non-overlapped matrix in a single object, by storing the ``owned'' rows first, followed by the off-processor ones. This arrangement allows to build the non-overlapped matrix as a ``subview'' of the overlapped one. The benefit is twofold: the memory footprint for the Jacobian is roughly halved, plus no copy is needed to transfer data for the local rows from the overlapped Jacobian to the non-overlapped Jacobian. This translated to a speedup of roughly 1.1 on CPUs and 2.1 on GPUs when analyzing finite element assembly performance relative to the old implementation.

\subsection{Boundary conditions} \label{subsec:impr-bcs}
\begin{figure}[htbp]
\centering
\includegraphics{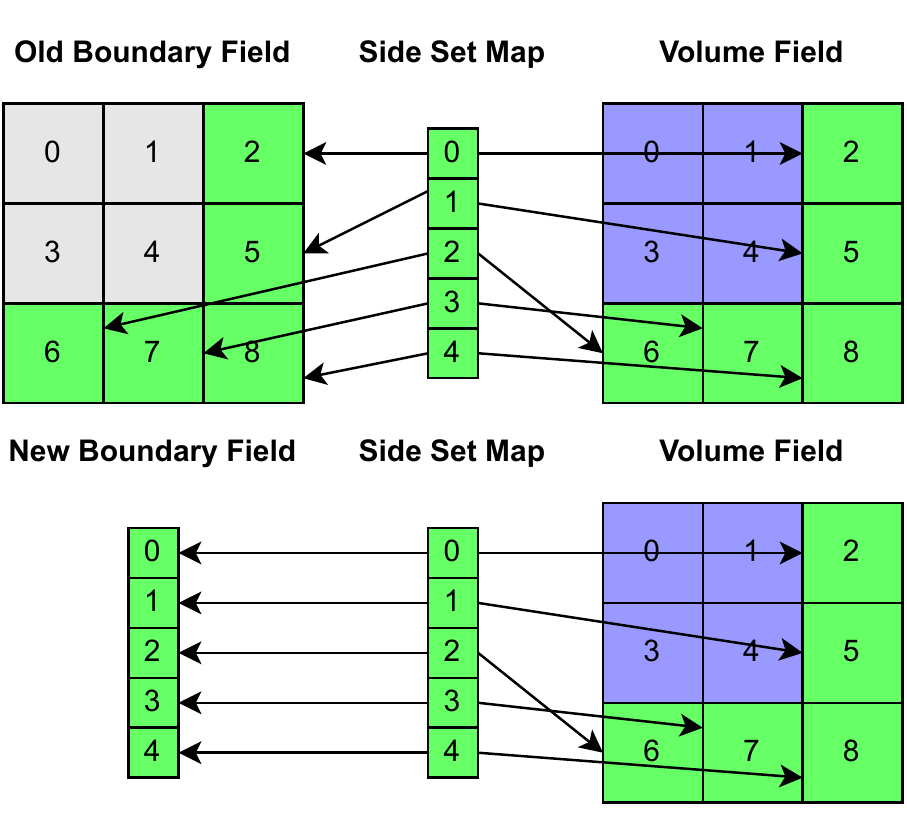}
\caption{Boundary data were originally stored as a volume field combined with a mapping data structure for accessing appropriate boundary cells. It is now aligned to match one-to-one with the side set map and can be read in a coalesced fashion on the device.}
\label{fig:sideset-layout}
\end{figure}

In order to achieve high performance on GPUs, fields corresponding to boundary data needed to be aligned for coalesced access on the device. The process is described in detail in \cite{carlson2020improvements} for a different problem and is summarized in Figure \ref{fig:sideset-layout}.
Originally, boundary data were stored using the same layout as a volume field with a data structure that contained a list of indices corresponding to cells that belong to the boundary. This effectively meant that each thread in a device block was loading data from non-consecutive locations in memory which is a highly inefficient access pattern for GPUs. By aligning boundary data to match the layout of the side set mapping data structure, all boundary fields are now read efficiently within a device kernel. Modifying the boundary data layouts had the additional benefit of significantly reducing memory usage for both CPU and GPU. A speedup of roughly 1.2 and 8.7 was achieved on CPUs and GPUs, respectively, when analyzing finite element assembly performance relative to the old implementation.

\subsection{Matrix dependent semicoarsening algebraic multigrid} \label{subsec:impr-MDSC-AMG}
Performance portable SA-AMG is provided by Trilinos/MueLu and is activated by using the Kokkos version of each component. This was extended to also include performance portable matrix dependent grid transfers for semicoarsening to complete the performance portable MDSC-AMG preconditioner introduced in Section~\ref{subsec:impl-linsol}.

The Kokkos version of MDSC (MDSC-Kokkos) uses Kokkos \lstinline{View} as temporary data structures while assembling the prolongation matrix. Kokkos \lstinline{parallel_for} is used to fill the contribution from each vertical line in parallel. A block tridiagonal system is assembled for each coarse layer in a vertical line on the fly. The system is then solved inline within the kernel using \lstinline{KokkosBatched} \lstinline{SerialLU} from Kokkos Kernels \cite{trott2021kokkos} where each thread performs an LU factorization without pivoting. The performance portable implementation is designed specifically as a batched direct solver for many small matrices and, in this case, additional optimizations are not needed. Once the solution is obtained, it is placed directly in the prolongation matrix.

Three performance portable point smoothers are provided by Trilinos/Ifpack2: MT Gauss-Seidel, Two-stage Gauss-Seidel, and Chebyshev. An autotuning framework using random search via Scikit-learn \cite{scikit-learn} is developed and used to determine the best smoother parameters for a small ALI test problem. We found that the performance portable smoothers did not outperform the serial line and point Gauss-Seidel smoothers in CPU simulations but Chebyshev smoothers performed the best in GPU simulations. Thus, the best CPU simulations continue to use the original line and point Gauss-Seidel smoothers while GPU simulations utilize the Chebyshev smoothers.

\subsection{Automated performance testing} \label{subsec:impr-test}
Performance tests are constructed as an extension of nightly regression tests. For example, a regression test might compute the steady-state solution of Equation~\eqref{eq:FOStokes} on a coarse Greenland mesh and compare computed surface velocities with known surface velocities. A performance test would perform the same calculation but also compare the end-to-end wall-clock time of the simulation to a specified value. Unfortunately, HPC clusters regularly exhibit large variations in performance causing performance tests to fail without any changes to the software. Thus, this method of performance testing is rarely used and changes in performance can go unnoticed for weeks or months.

A fundamental problem in maintaining performance tests is the ability to assess variations in performance on HPC systems. This requires a statistical approach to determine performance regressions and improvements. This is exemplified in \cite{hoefler2015scientific} where the authors provide methods of measuring and reporting performance on HPC systems. Performance regressions, or performance degradation in software execution, can occur through various mechanism, including changes in compilers, third party libraries, hardware, or software. As the number of developers of a scientific software stack grows, the likelihood of performance regressions increase. This is addressed by developing a framework that automatically collects performance metrics and applies a changepoint detection method to the data to detect changes in performance during nightly testing.

Changepoint detection is well researched in many fields \cite{aminikhanghahi2017survey,brodsky2016change,tartakovsky2014sequential,daly2020use} and is the process of finding abrupt variations in time series data. A changepoint detection method performs hypothesis testing between the null hypothesis, $H_0$, where no change occurs and the alternative hypothesis, $H_A$, where a change occurs. Given the performance metric time series,
\begin{equation}
X = \left\{x_1, x_2, \ldots, x_n\right\},
\end{equation}
where $n$ is the number of historical samples collected while testing for a performance metric, $x$, a subset of $X$ can be defined as,
\begin{equation}
X_i^j = \left\{x_i,  x_{i+1}, \ldots, x_j\right\},
\end{equation}
where $i$ and $j$ are the lower and upper limits of the time series, $X_i^j$. Two families of hypotheses are formulated as,
\begin{equation}
\begin{alignedat}{2}
H_0 &: f_1^{\nu-1} = f_{\nu}^n, \qquad \forall &&\nu \in \mathcal{K}, \\
H_A &: f_1^{\nu-1} \neq f_{\nu}^n, \qquad &&\nu \in \mathcal{K},
\end{alignedat}
\end{equation}
where $f_i^j$ is the probability density function $\forall x \in X_i^j$ and $\mathcal{K} = \left\{2, 3, \ldots, n\right\}$. This can be viewed as a generalized likelihood ratio test where $H_0$ states that all $x \in X$  belong to a single probability distribution while $H_A$ states that there exists some $\nu$ such that all $x \in X_1^{\nu-1}$ and $x \in X_{\nu}^n$ belong to two separate probability distributions, respectively \cite{hawkins2003changepoint}.

A two-sample $t$-test of $X_1^{\nu-1}$ and $X_{\nu}^n$ is performed to determine whether $\nu$ is a potential changepoint. In order to perform multiple hypothesis tests, the Bonferroni correction \cite{bonferroni1936teoria} is used to adjust the significance level by $\alpha / k$ where $\alpha$ is the desired significance level and $k = n-1$ is the number of tests. This correction is known to be overly conservative for large numbers of tests so only the largest changes in the time series are chosen. The pseudocode for this method is shown in Algorithm~\ref{alg:changepoint}.
\begin{algorithm}[htbp]
\caption{The single changepoint detection algorithm performs $k$ hypothesis tests on the time series $X_1^n$ using the significance level $\alpha$. The time series is assumed to be roughly normal. The result is a set of changepoints, $C$, and $t$-statistics, $S$. ArgSortDesc() performs an indirect sort from largest to smallest values and returns an array of indices that would sort the array. This is used to find the $k$ largest jumps.}
\label{alg:changepoint}
\begin{algorithmic}[1]
\Function{Changepoint}{$X_1^n, \alpha, k$}
  \State $C = []$
  \State $S = \{\}$
  \State $t^* = t^{1-\alpha/\left(2k\right)}_{n-2}$
  \State $\mathcal{K} = 1+\Call{ArgSortDesc}{\left|X_2^n - X_1^{n-1}\right|}$
  \For{$\nu \in \mathcal{K}_1^k$}
    \State $t = \Call{T-Test}{X_1^{\nu-1},X_\nu^n}$
    \If{$|t| > t^*$}
      \State $C \gets \left[C,\nu\right]$
      \State $S \gets S \cup \{\nu \to t\}$
    \EndIf
  \EndFor
  \Return{$C,S$}
\EndFunction
\end{algorithmic}
\end{algorithm}

A performance metric time series is likely to contain multiple changepoints. A sequential method is used to differentiate the time series based on previously identified changepoints. Once a changepoint is detected, the method disregards any data prior to the changepoint. This ensures that changepoints are not retroactively changed as new data are introduced.

Due to the large variation in HPC systems, the time series data may also contain outliers which can be identified as changepoints. Multiple methods are used to ensure that the changepoints are accurate in the presence of outliers:
\begin{enumerate}
\item In any single $t$-test, outliers are identified on both distributions using the median absolute deviation \cite{leys2013detecting} with a threshold comparable to three standard deviations. We remove at most 10\% of the total data.
\item A minimum number of consecutive detections, $m$, of the same changepoint are needed before confirming a changepoint. This helps dilute the influence of an outlier.
\item As the time series is traversed sequentially, the sample size or ``lookback window" for each test is limited by $w$ observations. This helps avoid hypersensitivity where the smallest change in the time series becomes significant when the sample size is too large.
\end{enumerate}

Algorithm~\ref{alg:changepoint} along with its modifications for multiple changepoint detection is implemented in python and executed during nightly performance testing. The log-transformed values are used because a log-normal distribution seems to fit the data slightly better than a normal distribution. A significance level of $\alpha=0.005$ is chosen to ensure confidence and only the $k=10$ largest changes are considered. $m=3$ consecutive detections are needed to confirm a changepoint and a lookback window of $w=30$ is chosen. This typically means that a minimum of three days are needed to detect a changepoint but this depends on data variability. Daily results are reported on an automated Jupyter notebook and posted online (see \url{https://sandialabs.github.io/ali-perf-data/}), and performance regressions are reported through an automated email report.

Performance regressions and improvements are quantified by utilizing changepoints to define subsets within the time series. Given a changepoint and two subsets, a 99\% confidence interval (CI) for the difference in mean on log-transformed values is computed using a $t$-distribution. When transformed back, a relative performance ratio (speedup or slowdown) is given for the regression or improvement. Equation~\eqref{eq:mean-transform} shows an example of how the relative performance is computed,
\begin{equation} \label{eq:mean-transform}
\frac{\overline{\log\left(X_{\nu}^n\right)}}{\overline{\log\left(X_1^{\nu-1}\right)}} = \exp\left(\overline{\log\left(X_{\nu}^n\right)} - \overline{\log\left(X_1^{\nu-1}\right)}\right),
\end{equation}
where the overline represents an arithmetic mean. A similar technique is also used in Sections~\ref{subsec:res-impr} and~\ref{subsec:res-weak} to compute speedups, proportions and efficiencies with 99\% confidence intervals.

For performance comparisons, the difference in mean on log-transformed data between two performance tests needs to be established. A paired $t$-test is performed by taking the difference between the log-transformed data for the two tests where the dates intersect. The changepoint detection method is used on this data to identify subsets and a 99\% confidence interval (CI) for the difference in mean on log-transformed data is computed on each subset. This is also computed as a relative performance ratio when transformed back.

\section{Numerical results} \label{sec:res}
In this section, the performance of MALI and standalone ALI is analyzed on two variable resolution Greenland ice-sheet meshes and a series of increasing higher resolution Antarctic ice-sheet meshes. In the first Greenland case, MALI is compared with and without the features described in Sections~\ref{subsec:impr-mem} and~\ref{subsec:impr-MDSC-AMG}, and performance improvements are shown across all HPC architectures. In the second case, a weak scalability study of Antarctica shows that simulations perform best when utilizing the GPUs on modern HPC systems. In the last Greenland case, several examples are given on how the changepoint detection method described in Section~\ref{subsec:impr-test} is used to identify performance regressions, improvements and differences in algorithm performance. What follows is a brief description of the experimental setup.

The MALI code base consists of three open-source software projects that are continuously updated through github repositories and tested nightly for performance and correctness on HPC machines. Table~\ref{tab:res-MALI-software} shows where the three projects currently exist and the commit ids used for the performance experiments to follow.
\begin{table}[htbp]
\caption{MALI software repositories}
\label{tab:res-MALI-software}
\begin{tabular}{p{2cm}p{7.5cm}p{2.5cm}p{2.5cm}}
\hline\noalign{\smallskip}
Software & Repository & Git Branch & Commit Id\\
\noalign{\smallskip}\hline\noalign{\smallskip}
MPAS & \url{https://github.com/MALI-Dev/E3SM} & develop & d6309858d9\\
Albany & \url{https://github.com/sandialabs/Albany} & master & 9d292d8f5\\
Trilinos & \url{https://github.com/trilinos/Trilinos} & develop & 155e45e86c2\\
\noalign{\smallskip}\hline\noalign{\smallskip}
\end{tabular}
\end{table}
The code is compiled with the Kokkos \lstinline{Serial} execution space for CPU-only simulations and \lstinline{Cuda} for simulations utilizing GPUs. CPU-only simulations are executed with MPI ranks mapped to cores while GPU simulations are executed with MPI ranks mapped to GPUs. In all experiments, CUDA-Aware MPI is turned off and \lstinline{CUDA_LAUNCH_BLOCKING} is turned on.

The simulations are executed on the four architectures provided on the Cori and Summit supercomputers. A summary of each testing environment is provided in Table~\ref{tab:res-clusters}.
\begin{table}[htbp]
\caption{MALI simulations are executed on the three platforms or four architectures given below. A limited number of cores are utilized on some systems in order to keep a core idle for system operations. On Summit, an MPI-only simulation using only the CPU is tested along with an MPI+GPU simulation.}
\label{tab:res-clusters}
\begin{tabular}{p{3.0cm}p{3.9cm}p{3.9cm}p{3.9cm}}
\hline\noalign{\smallskip}
Name & Cori (HSW) & Cori (KNL) & Summit (PWR9,V100) \\
\noalign{\smallskip}\hline\noalign{\smallskip}
CPU & Intel Xeon E5-2698 v3\newline Haswell & Intel Xeon Phi 7250\newline Knights Landing & IBM POWER9 \\
Number of Cores & 16 & 68 & 22 \\
GPU & None & None & NVIDIA Tesla V100 \\
Node Arch. & 2 CPUs & 1 CPU & 2 CPUs + 6 GPUs \\
Memory per Node & 125 GiB & 94 GiB & 604 GiB +\newline 15.7 GiB/GPU \\
CPU Compiler & Intel 19.0.3.199 & Intel 19.0.3.199 & gcc 9.1.0 \\
GPU Compiler & None & None & nvcc 11.0.3 \\
MPI & cray-mpich 7.7.10 & cray-mpich 7.7.10 & spectrum-mpi 10.4.0.3 \\
Node Config. & 32 MPI & 64 MPI & PWR9: 42 MPI\newline V100: 6 MPI \\
\noalign{\smallskip}\hline\noalign{\smallskip}
\end{tabular}
\end{table}
Wall-clock time is captured by using MPAS and Trilinos/Teuchos timers to obtain an average time across MPI processes. The relevant timers and their descriptions are shown in Table~\ref{tab:res-timers}.
\begin{table}[htbp]
\caption{The MALI timers described below are used to collect the average wall-clock time across MPI processes.}
\label{tab:res-timers}
\begin{tabular}{p{3.5cm}p{12.0cm}}
\hline\noalign{\smallskip}
Timer & Description \\
\noalign{\smallskip}\hline\noalign{\smallskip}
Total Time & Total simulation time reported by MALI \\
Total Solve & Total nonlinear solve time reported by ALI \\
Total Fill & Total finite element assembly time starting from the ice velocity import and ending with the export of the residual and Jacobian \\
Preconditioner\newline Construction & Total time constructing the MDSC-AMG preconditioner  \\
Linear Solve & Total time in the linear solver including the application of the preconditioner \\
\noalign{\smallskip}\hline\noalign{\smallskip}
\end{tabular}
\end{table}

\subsection{MALI Greenland ice-sheet 1-to-10 kilometer variable resolution case} \label{subsec:res-impr}
\begin{figure}[htbp]
\centering
\begin{subfigure}{0.48\textwidth}
\centering
\includegraphics[width=\textwidth]{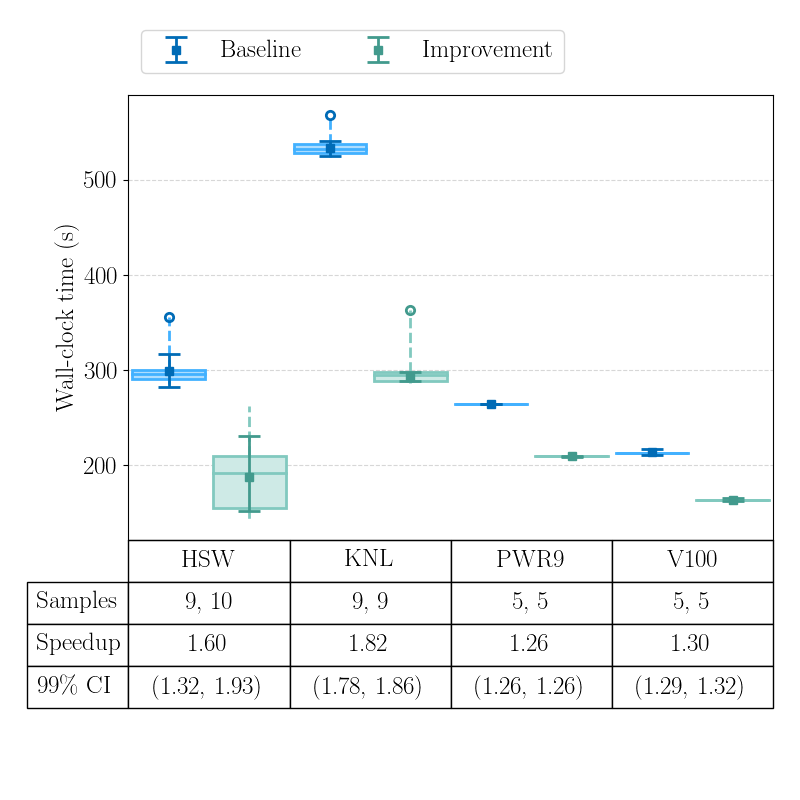}
\caption{Total Time}
\label{fig:res-impr-total}
\end{subfigure}
\hfill
\begin{subfigure}{0.48\textwidth}
\centering
\includegraphics[width=\textwidth]{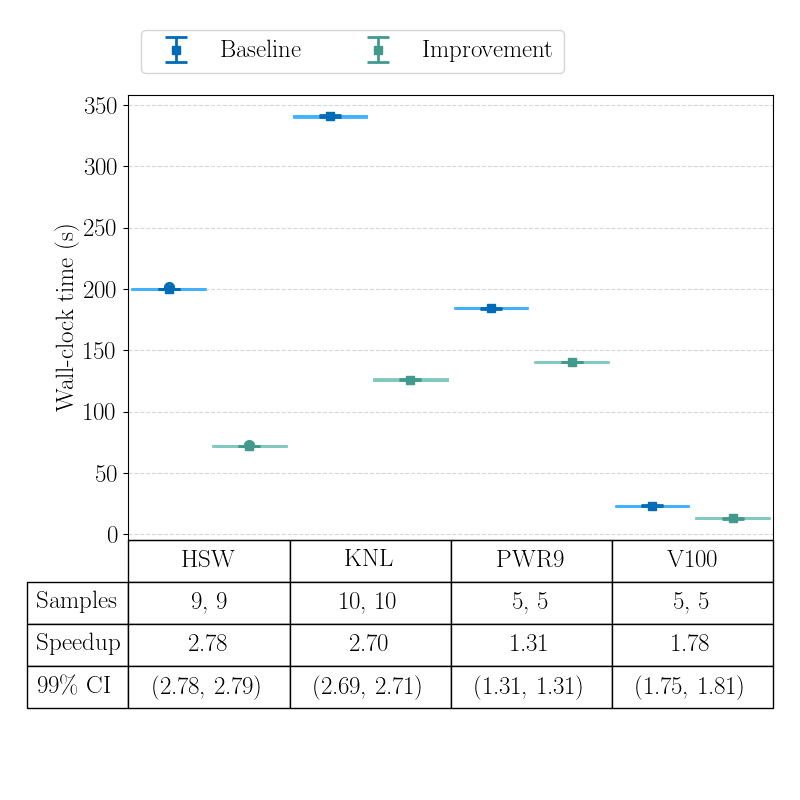}
\caption{Total Fill}
\label{fig:res-impr-fill}
\end{subfigure}
\vskip\baselineskip
\begin{subfigure}{0.48\textwidth}
\centering
\includegraphics[width=\textwidth]{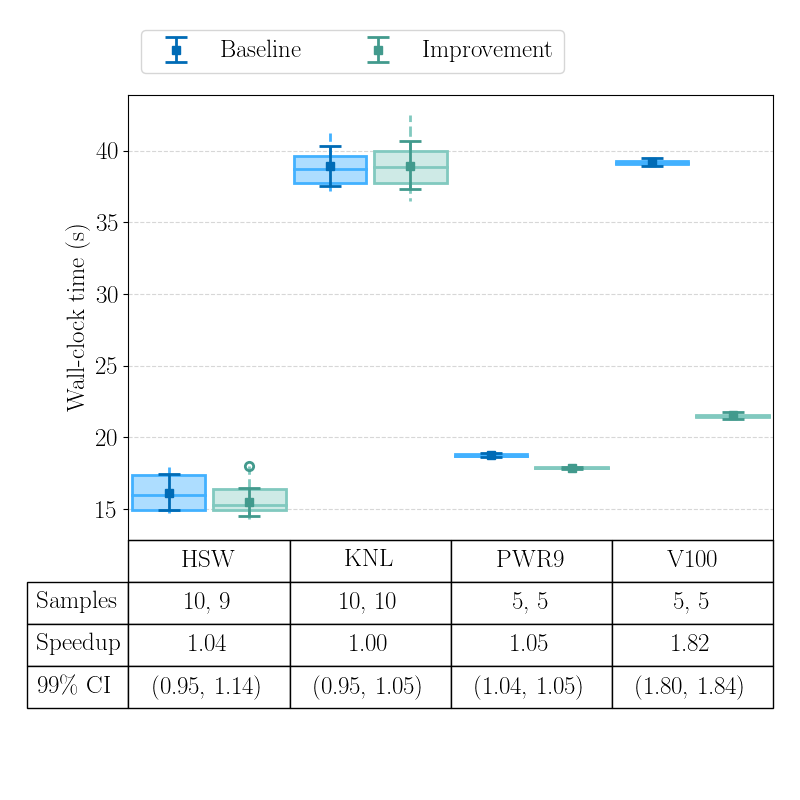}
\caption{Preconditioner Construction}
\label{fig:res-impr-prec}
\end{subfigure}
\hfill
\begin{subfigure}{0.48\textwidth}
\centering
\includegraphics[width=\textwidth]{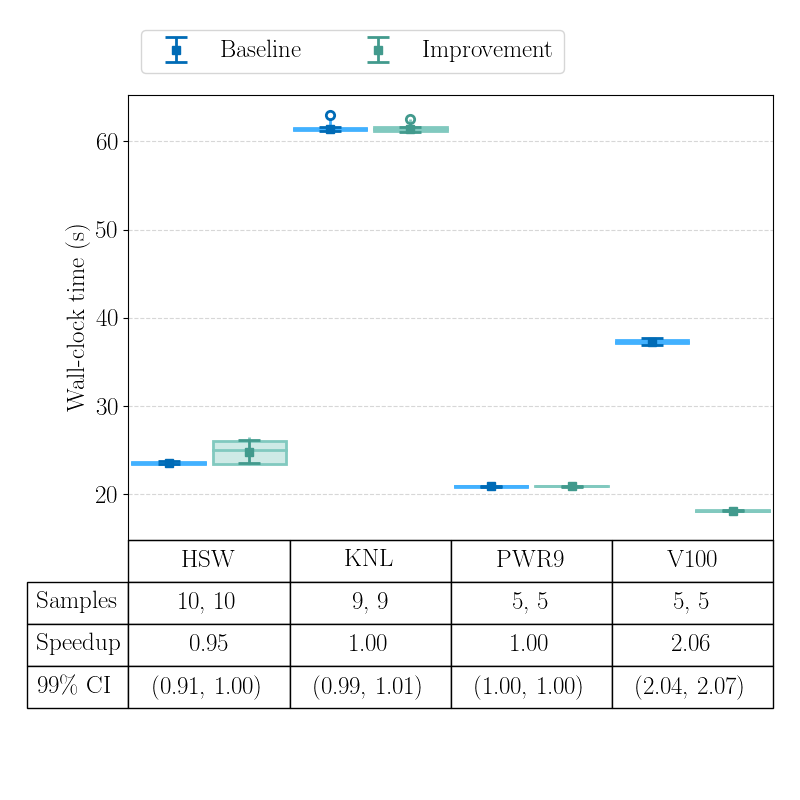}
\caption{Linear Solve}
\label{fig:res-impr-linsol}
\end{subfigure}
\caption{The MALI Greenland ice-sheet 1-to-10 km variable resolution simulation is executed multiple times on four architectures (Cori: HSW, KNL; Summit: PWR9, V100) and two cases in order to capture improvements across four timers. Architectures, timers and cases are defined in Table~\ref{tab:res-clusters},~\ref{tab:res-timers} and~\ref{tab:res-impr-cases}, respectively. The lower/upper quartiles are shown along with the median in a box plot while a dashed line is used to show the full data range. The sample size is trimmed using the methods discussed in Section~\ref{subsec:impr-test} and outliers are shown as circles. The trimmed sample sizes for the baseline and improvement are given in the table as a pair and a mean error bar is plotted. The speedup from the improvement relative to the baseline is also given along with a confidence interval (CI). CIs are reported as (LL, UL) where LL is the lower limit and UL is the upper limit.}
\label{fig:res-impr}
\end{figure}

Here we consider a variable-resolution grid of the Greenland ice-sheet, which is finer in regions with a more complex flow structure, i.e. close to the margin and in regions where the observed surface velocity is higher. The 2D grid with cell spacing ranging from \num{1}km to \num{10}km is extruded in the vertical direction using \num{10} layers of variable thickness (thinner at the bed). The basal sliding condition and temperature are pre-computed using an initialization approach that matches the surface velocity observation while satisfying the first-order velocity equations coupled to a steady state enthalpy equation \cite{perego2014optimal,heinlein2022frosch}. In this case, MALI is used to perform an initial state calculation and a single time step, leading to two nonlinear solves using ALI. The temperature is held fixed during the time step. The nonlinear and linear solver tolerances are \num{1e-5} and \num{1e-8}, respectively. Simulations are compared with and without the features described in Sections~\ref{subsec:impr-mem} and~\ref{subsec:impr-MDSC-AMG}. The cases are given in Table~\ref{tab:res-impr-cases}.
\begin{table}[htbp]
\caption{The MALI Greenland ice-sheet 1-to-10 km variable resolution simulation is executed with and without the specific features described below.}
\label{tab:res-impr-cases}
\begin{tabular}{p{3.5cm}p{12.0cm}}
\hline\noalign{\smallskip}
Case Name & Description \\
\noalign{\smallskip}\hline\noalign{\smallskip}
Baseline & Finite element assembly without memoization, serial MDSC, default smoother settings \\
Improvement & Finite element assembly with memoization, MDSC-Kokkos, optimal smoothers found through autotuning \\
\noalign{\smallskip}\hline\noalign{\smallskip}
\end{tabular}
\end{table}
Multiple samples are collected for each case using the same allocation and a mean error bar is computed using the method described in Section~\ref{subsec:impr-test}. A two-sample $t$-test of the mean difference of the log is also performed to ensure differences are statistically significant. This is then used to compute a confidence interval for the speedup of the improvement relative to the baseline. The results for each timer are shown in Figure~\ref{fig:res-impr}.

The \textbf{Total Fill} timer in Figure~\ref{fig:res-impr-fill} is used to measure the improvement from memoization. The variation from this timer is small and the performance improvement is clear across all architectures. Larger improvements are seen on Cori. More variation is seen from the \textbf{Preconditioner Construction} timer in Figure~\ref{fig:res-impr-prec} which is used to measure the improvement from MDSC-Kokkos. In this case, the speedup on Cori is not statistically significant but the speedup on POWER9 shows that there may be some benefit on CPU architectures. The speedup on V100 GPUs is larger and more significant. Lastly, the \textbf{Linear Solve} timer in Figure~\ref{fig:res-impr-linsol} is used to measure the improvement from tuning the GPU preconditioner. There is some variation and performance loss seen in the linear solve on Haswell CPUs but it is not very significant and the slowdown is not seen on the other CPU architectures. Again, there is a statistically significant speedup on V100 GPUs.

The performance of the linear solver is highlighted in Table~\ref{tab:res-impr-iters}.
\begin{table}[htbp]
\caption{The MALI Greenland ice-sheet 1-to-10 km variable resolution simulation is executed multiple times on four architectures (Cori: HSW, KNL; Summit: PWR9, V100) and two cases in order to capture improvements in the linear solve. Architectures and cases are defined in Table~\ref{tab:res-clusters} and~\ref{tab:res-impr-cases}, respectively. The table below shows the average number of linear iterations and the total linear solve time across 26 nonlinear iterations (2 nonlinear solves) for all cases. A 99\% confidence interval is reported (when statistically significant) as (LL, UL) where LL is the lower limit and UL is the upper limit.}
\label{tab:res-impr-iters}
\begin{tabular}{p{2.0cm}p{3.0cm}p{4.0cm}p{4.0cm}}
\hline\noalign{\smallskip}
 & Case & Avg. Lin. Its. & Linear Solve Time (s) \\
\noalign{\smallskip}\hline\noalign{\smallskip}
HSW & Baseline    & 12.0 & 23.6 (23.4, 23.8)\\
    & Improvement & 12.0 & 24.8 (23.5, 26.1)\\
\noalign{\smallskip}\hline\noalign{\smallskip}
KNL & Baseline    & 12.0 & 61.4 (61.1, 61.6)\\
    & Improvement & 12.0 & 61.4 (61.1, 61.7)\\
\noalign{\smallskip}\hline\noalign{\smallskip}
PWR9 & Baseline    & 11.7 & 20.9 (20.8, 21.0)\\
     & Improvement & 11.7 & 20.9 \\
\noalign{\smallskip}\hline\noalign{\smallskip}
V100 & Baseline    & 43.9 (43.5, 44.2) & 37.3 (36.9, 37.7)\\
     & Improvement & 48.3 (48.2, 48.5) & 18.1 (18.1, 18.2)\\
\noalign{\smallskip}\hline\noalign{\smallskip}
\end{tabular}
\end{table}

\begin{figure}[htbp]
\centering
\includegraphics[width=\textwidth]{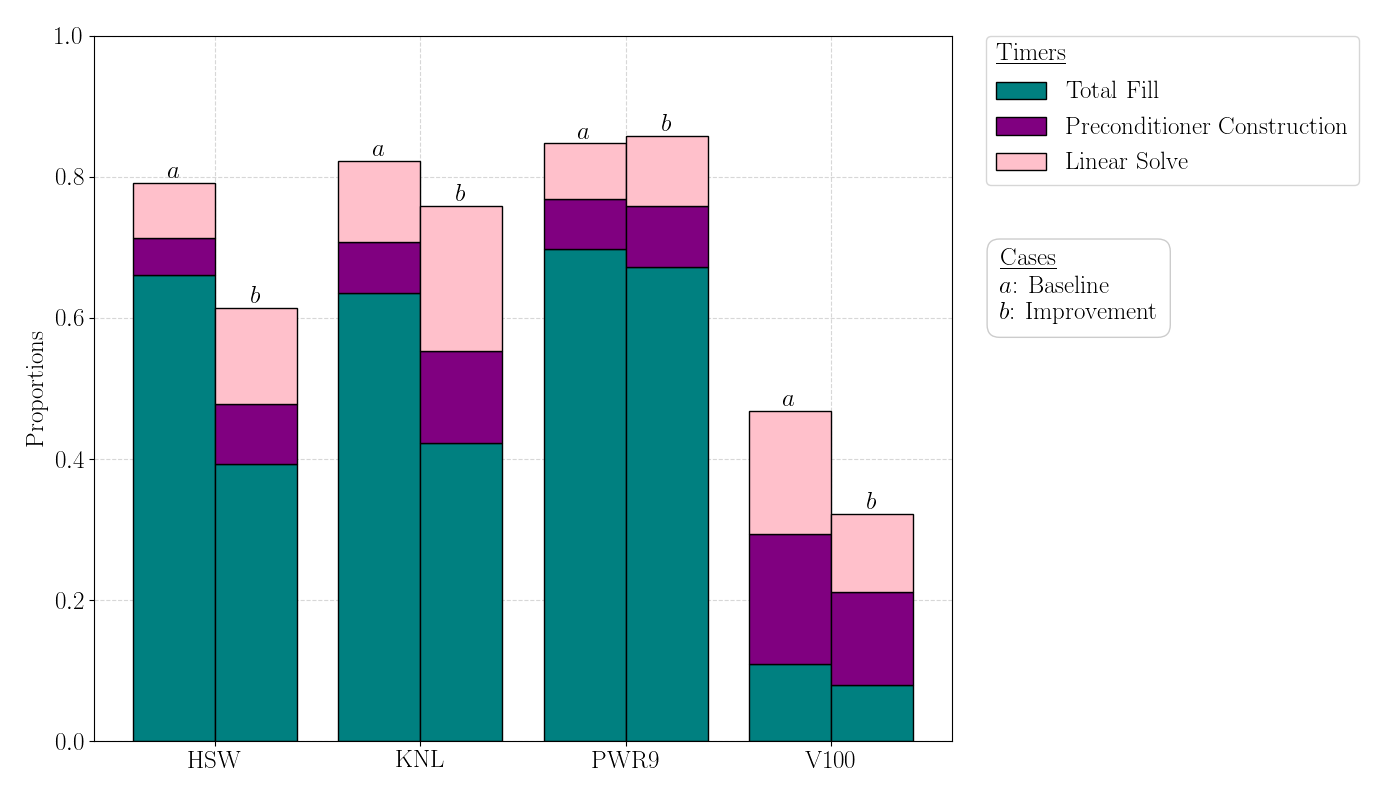}
\caption{The MALI Greenland ice-sheet 1-to-10 km variable resolution simulation is executed multiple times on four architectures (Cori: HSW, KNL; Summit: PWR9, V100) and two cases. Architectures, timers and cases are defined in Table~\ref{tab:res-clusters},~\ref{tab:res-timers} and~\ref{tab:res-impr-cases}, respectively. This plot shows the ratio of each timer compared to \textbf{Total Time}.}
\label{fig:res-impr-prop}
\end{figure}

Figure~\ref{fig:res-impr-total} shows the overall performance improvement in MALI from the addition of memoization, MDSC-Kokkos and GPU preconditioner tuning. There is a statistically significant performance improvement across all architectures despite the large variation on Cori. Lastly, Figure~\ref{fig:res-impr-prop} shows the proportions of total wall-clock for each architecture, case and timer.
The plot shows that on CPU platforms, \textbf{Total Fill} remains a significant portion of total runtime. On GPU platforms, the finite element assembly is much less significant when compared to the linear solver and a large portion of runtime is in other portions of the code. This includes the initial setup of the data structures which only executes once but is significantly more expensive when compared to CPU platforms.

\subsection{ALI Antarctica ice-sheet weak scalability study} \label{subsec:res-weak}
The second case focuses on solving the first-order velocity equations in a weak scalability of ALI on a series of structured Antarctica ice-sheet meshes. This case has been used in a number of other papers including \cite{tezaur2015scalability, tuminaro2016matrix, heinlein2022frosch} where more detailed descriptions are given. In this case, the focus will be on how well the CPU+GPU solver performs over the CPU-only version. The five meshes vary in resolution from 1 to 16 kilometers and quadrilateral element counts vary from 51,087 to 13,413,740. The mesh is extruded by 20 layers during the setup phase, the equation is solved using the methods described in Section~\ref{sec:impl}, and the mean value of the final solution is compared to a previously tested value using a relative tolerance of \num{1.0e-5} to ensure the results remain consistent across runs and architectures. The basal sliding coefficient is predetermined using deterministic inversion from observed surface velocities \cite{perego2014optimal} and a realistic temperature field is provided. Table~\ref{tab:res-weak-dofs} shows the number of compute nodes allocated and the total degrees of freedom for each case.
\begin{table}[htbp]
\centering
\caption{A series of increasingly higher resolution Antarctic ice-sheet simulations are executed in a weak scalability study on four architectures. The table below shows the computing resources and degrees of freedom (DOF) associated with each case.}
\label{tab:res-weak-dofs}
\begin{tabular}{p{2.5cm}p{2.5cm}p{2.5cm}}
\hline\noalign{\smallskip}
Resolution & Nodes & DOF \\
\noalign{\smallskip}\hline\noalign{\smallskip}
16km & 1   & \num{2.20e6} \\ 
 8km & 4   & \num{8.83e6} \\ 
 4km & 16  & \num{3.53e7} \\ 
 2km & 64  & \num{1.41e8} \\ 
 1km & 256 & \num{5.66e8} \\ 
\noalign{\smallskip}\hline\noalign{\smallskip}
\end{tabular}
\end{table}

\begin{figure}[htbp]
\centering
\begin{subfigure}{0.48\textwidth}
\centering
\includegraphics[width=\textwidth]{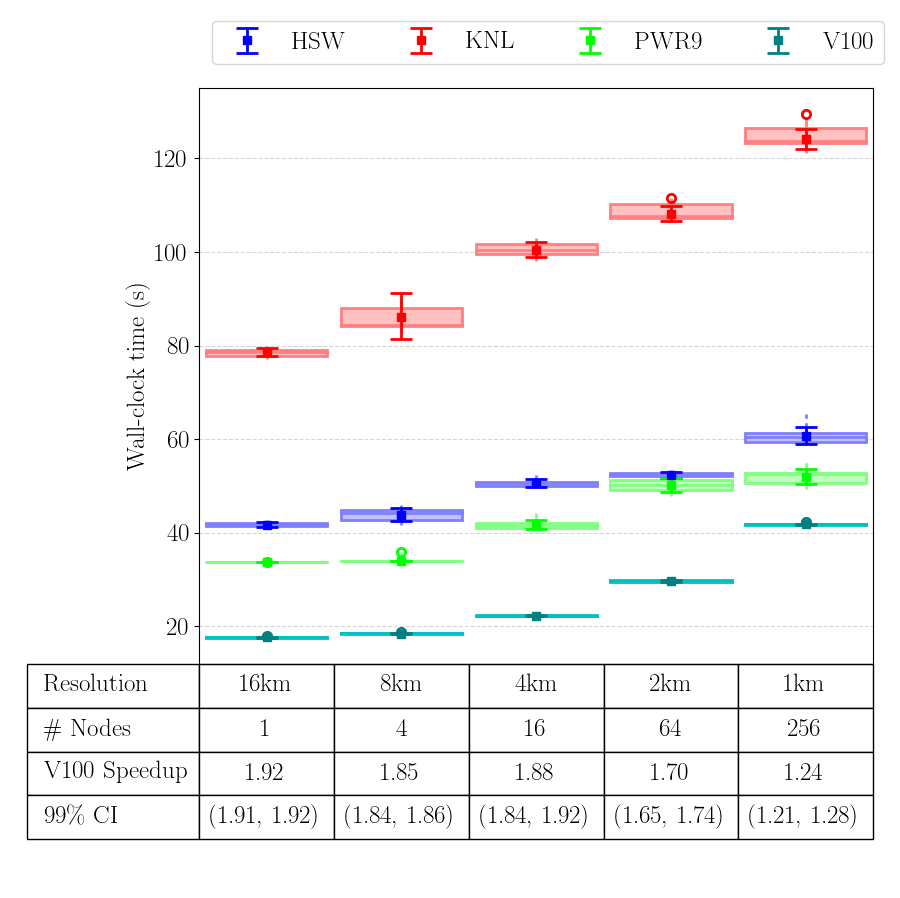}
\caption{Total Solve}
\label{fig:res-weak-total}
\end{subfigure}
\hfill
\begin{subfigure}{0.48\textwidth}
\centering
\includegraphics[width=\textwidth]{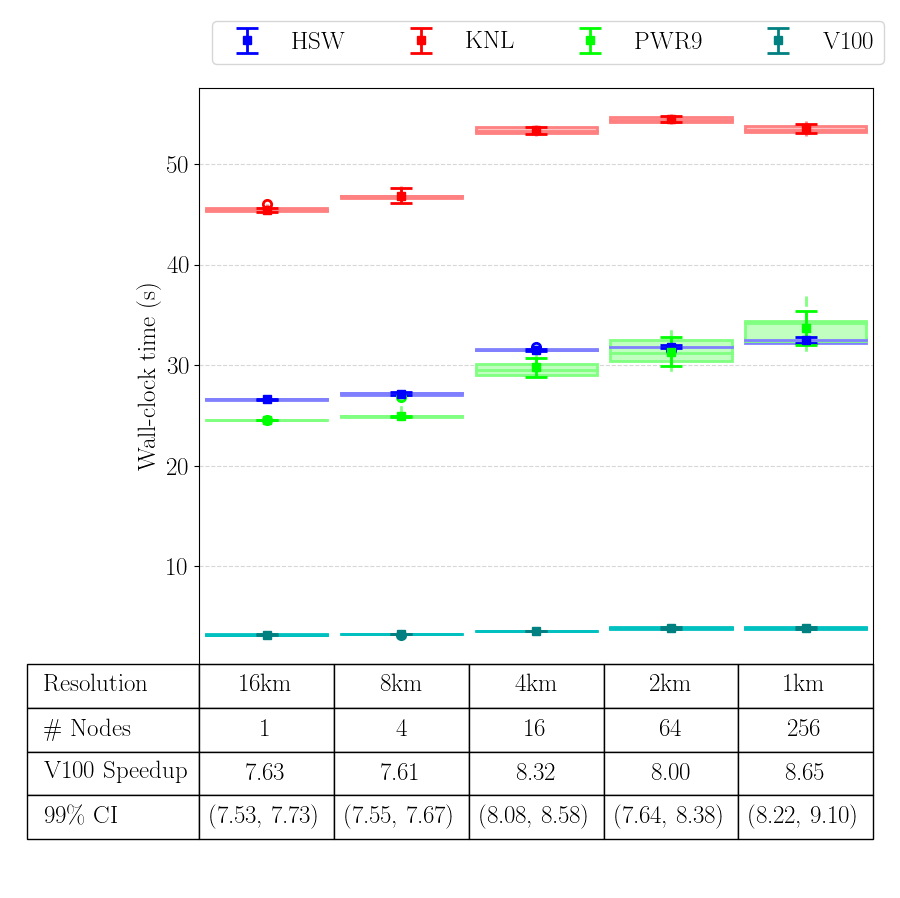}
\caption{Total Fill}
\label{fig:res-weak-fill}
\end{subfigure}
\vskip\baselineskip
\begin{subfigure}{0.48\textwidth}
\centering
\includegraphics[width=\textwidth]{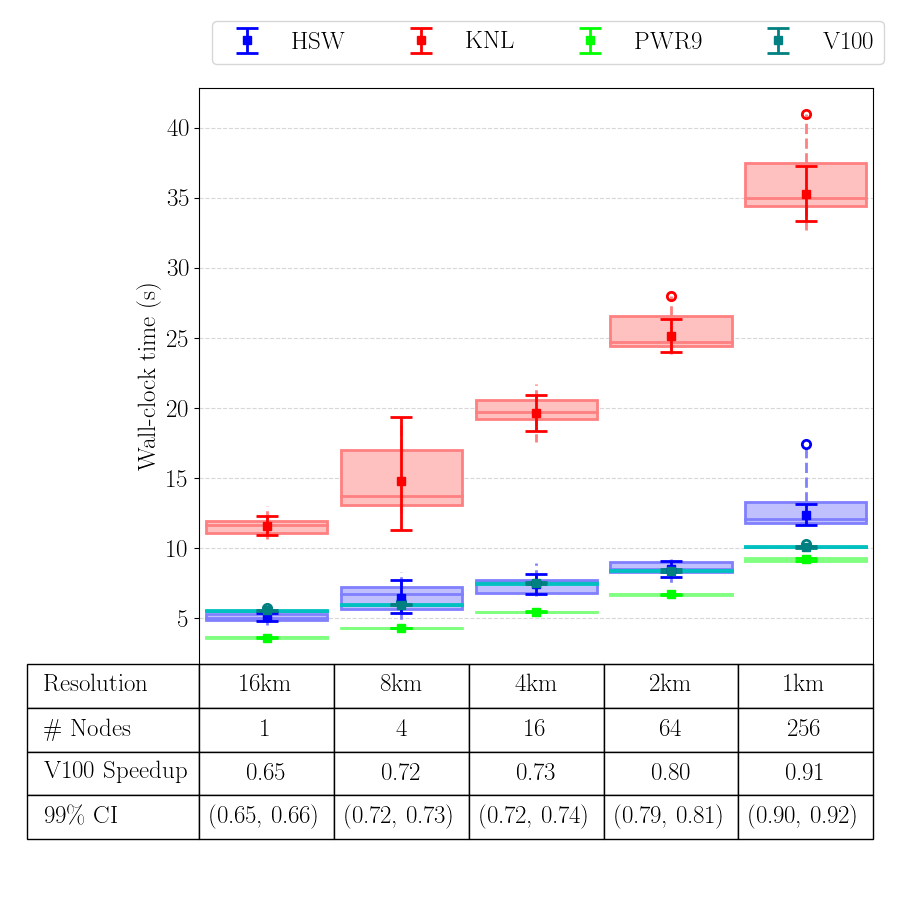}
\caption{Preconditioner Construction}
\label{fig:res-weak-prec}
\end{subfigure}
\hfill
\begin{subfigure}{0.48\textwidth}
\centering
\includegraphics[width=\textwidth]{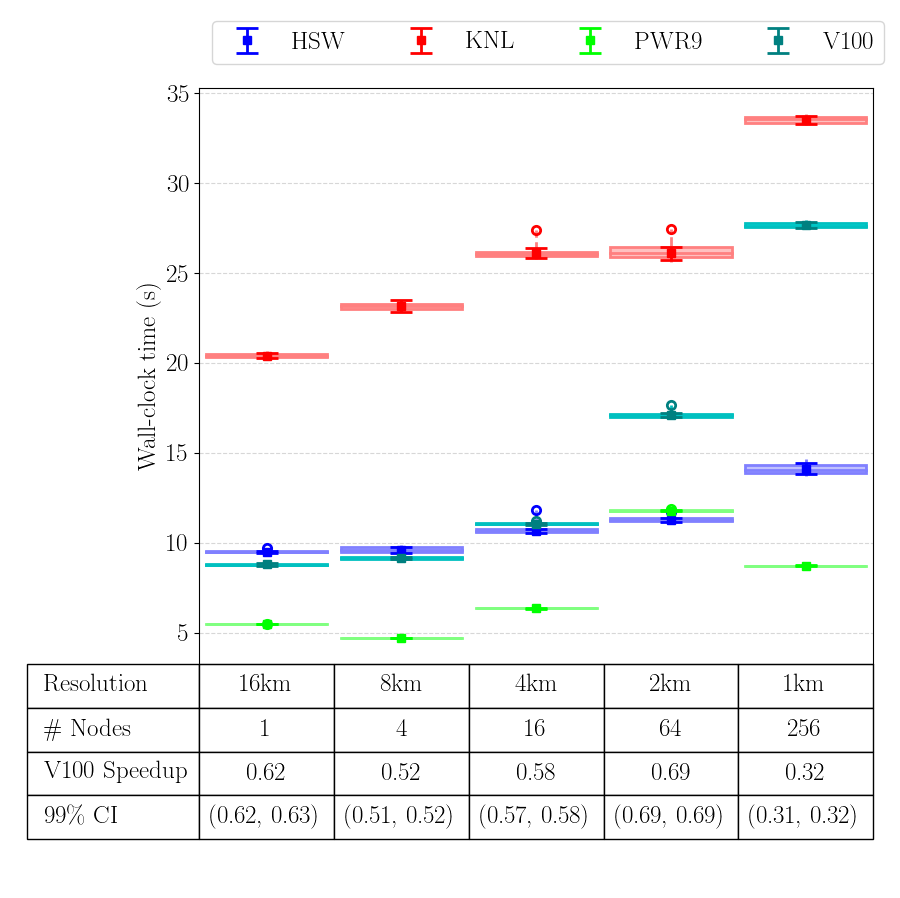}
\caption{Linear Solve}
\label{fig:res-weak-linsol}
\end{subfigure}
\caption{A series of increasingly higher resolution Antarctic ice-sheet simulations are executed in a weak scalability study on four architectures (Cori: HSW, KNL; Summit: PWR9, V100). Four timers are captured. These are defined in Table~\ref{tab:res-timers}. The mean error bar, median and lower/upper quartiles of each case are given along with outliers shown as circles. The speedup from the V100 CPU-GPU simulation relative to the POWER9 CPU-only simulation is also shown along with a confidence interval (CI). CIs are reported as (LL, UL) where LL is the lower limit and UL is the upper limit.}
\label{fig:res-weak}
\end{figure}

In this case, standalone ALI is used to perform a single nonlinear solve where the tolerances for the nonlinear and linear solvers are set to \num{1e-5} and \num{1e-6}, respectively. Simulations are executed with all improvements described in Section~\ref{sec:impr}. Similar to the Greenland case in Section~\ref{subsec:res-impr}, multiple samples are collected for each case using the same allocation and a mean error bar is computed. Two-sample $t$-tests of the mean difference of the log between the PWR9 and V100 cases are also performed and a 99\% confidence interval for the speedup of the GPU relative to the CPU-only simulation is given in Figure~\ref{fig:res-weak}.
The first notable result is that the \textbf{Total Fill} is around 8 times faster when utilizing the V100 GPUs. The same cannot be said about the \textbf{Preconditioner Construction} and \textbf{Linear Solve} where the performance is worse. Despite this performance loss, \textbf{Total Solve} is faster when utilizing the GPU. It is important to note that the PWR9, CPU-only \textbf{Linear Solve} performed exceptionally well compared to the other architectures.

Weak scaling is used to determine how well a code is able to maintain the same wall-clock time when simulating larger problems with a proportionally larger amount of resources. In this case, the problem size is not exactly proportional to the resource size so the following formula is used to compute the weak scaling efficiency in terms of percentages,
\begin{equation}
\eta = \frac{(t_1/N_1)/(t_n/N_n)}{n} \times 100 \%, \qquad 0 < \eta < 100\%,
\end{equation}
where $t$ is the wall-clock time, $n$ is the number of compute nodes, $N$ is the number of degrees of freedom, and larger values are better. Confidence intervals for the efficiency are computed by using the same mean difference of the log between the single compute node case and the 256 node case. The results are shown in Table~\ref{tab:res-weak-eff}.
\begin{table}[htbp]
\caption{A series of increasingly higher resolution Antarctic ice-sheet simulations are executed in a weak scalability study on up to 256 compute nodes on Cori and Summit. Four architectures (Cori: HSW, KNL; Summit: PWR9, V100) are tested and four timers are captured as defined in Table~\ref{tab:res-timers}. A weak scalability efficiency is computed for each case where one compute node is used as the reference. Larger values are better. A 99\% confidence interval is reported as (LL, UL) where LL is the lower limit and UL is the upper limit.}
\label{tab:res-weak-eff}
\begin{tabular}{p{2.0cm}p{3.0cm}p{3.0cm}p{3.0cm}p{3.0cm}}
\hline\noalign{\smallskip}
 & Total Solve & Total Fill & Preconditioner\newline Construction & Linear Solve\\
\noalign{\smallskip}\hline\noalign{\smallskip}
HSW & 68.9\% (67.0, 70.9) & 82.2\% (81.5, 82.9) & 41.2\% (38.2, 44.5) & 67.5\% (66.2, 68.8)\\
KNL & 63.5\% (62.3, 64.6) & 85.3\% (84.5, 86.0) & 33.0\% (30.8, 35.5) & 61.1\% (60.6, 61.6)\\
PWR9 & 65.1\% (63.3, 66.9) & 73.1\% (70.0, 76.4) & 39.5\% (39.0, 40.0) & 63.0\% (62.9, 63.1)\\
V100 & 42.2\% (42.0, 42.4) & 82.9\% (80.5, 85.4) & 55.2\% (54.7, 55.8) & 31.9\% (31.6, 32.2)\\
\noalign{\smallskip}\hline\noalign{\smallskip}
\end{tabular}
\end{table}
In this study, the Haswell CPU performed the best when looking at the \textbf{Total Solve} while the CPU+GPU case performed the worst. The \textbf{Total Fill} performed well across all architectures and there's a noticeable improvement on the GPU compared to previous studies \cite{watkins2020study}. In \textbf{Preconditioner Construction}, the CPU+GPU case performed best. The main cause for poor scaling on GPU platforms is visible in the \textbf{Linear Solve}. This can be explained by looking at the performance of the linear solver in Table~\ref{tab:res-weak-iters}.
\begin{table}[htbp]
\caption{A series of increasingly higher resolution Antarctic ice-sheet simulations are executed in a weak scalability study on four architectures (Cori: HSW, KNL; Summit: PWR9, V100). The table below shows the total number of nonlinear iterations (1 nonlinear solve), the average number of linear iterations per nonlinear iteration for all cases and the total linear solve time. A 99\% confidence interval is reported (when statistically significant) as (LL, UL) where LL is the lower limit and UL is the upper limit.}
\label{tab:res-weak-iters}
\begin{tabular}{p{1.5cm}p{2.0cm}p{1.5cm}p{2.0cm}p{3.5cm}p{3.0cm}}
\hline\noalign{\smallskip}
 & Resolution & Nodes & Nlin. Its. & Avg. Lin. Its. & Linear Solve Time (s) \\
\noalign{\smallskip}\hline\noalign{\smallskip}
HSW & 16km & 1   & 8 & 16.5 & 9.5s  (9.4, 9.5)\\
    &  8km & 4   & 8 & 15.5 & 9.6s  (9.4, 9.8)\\
    &  4km & 16  & 9 & 15.2 & 10.7s (10.6, 10.8)\\
    &  2km & 64  & 9 & 15.1 & 11.3s (11.2, 11.4)\\
    &  1km & 256 & 9 & 17.9 & 14.1s (13.8, 14.4)\\
\noalign{\smallskip}\hline\noalign{\smallskip}
KNL & 16km & 1   & 8 & 16.2 & 20.4s (20.3, 20.5)\\
    &  8km & 4   & 8 & 15.6 & 23.1s (22.8, 23.5)\\
    &  4km & 16  & 9 & 15.2 & 26.1s (25.8, 26.4)\\
    &  2km & 64  & 9 & 14.6 & 26.1s (25.7, 26.4)\\
    &  1km & 256 & 9 & 18.2 & 33.5s (33.3, 33.7)\\
\noalign{\smallskip}\hline\noalign{\smallskip}
PWR9 & 16km & 1   & 8 & 16.1 & 5.5s\\
     &  8km & 4   & 8 & 13.1 & 4.7s\\
     &  4km & 16  & 9 & 14.8 & 6.4s (6.3, 6.4)\\
     &  2km & 64  & 9 & 24.0 & 11.8s\\
     &  1km & 256 & 9 & 17.7 & 8.7s\\
\noalign{\smallskip}\hline\noalign{\smallskip}
V100 & 16km & 1   & 8 & 88.7 (88.6, 88.8) & 8.8s (8.7, 8.9)\\
     &  8km & 4   & 8 & 87.2 (87.1, 87.3) & 9.1s (9.1, 9.2)\\
     &  4km & 16  & 9 & 89.6 (89.6, 89.7) & 11.1s (11.0, 11.1)\\
     &  2km & 64  & 9 & 131.4 (131.2, 131.6) & 17.1s (17.0, 17.2)\\
     &  1km & 256 & 9 & 194.2 (193.6, 194.8) & 27.7s (27.5, 27.8)\\
\noalign{\smallskip}\hline\noalign{\smallskip}
\end{tabular}
\end{table}
The average number of linear iterations from the GPU linear solve is much larger (reaching the maximum iteration constraint) at 256 compute nodes which contributed to worse scaling. The PWR9, CPU-only case also has much smaller linear solve times compared to the other architectures.

\begin{figure}[htbp]
\centering
\includegraphics[width=\textwidth]{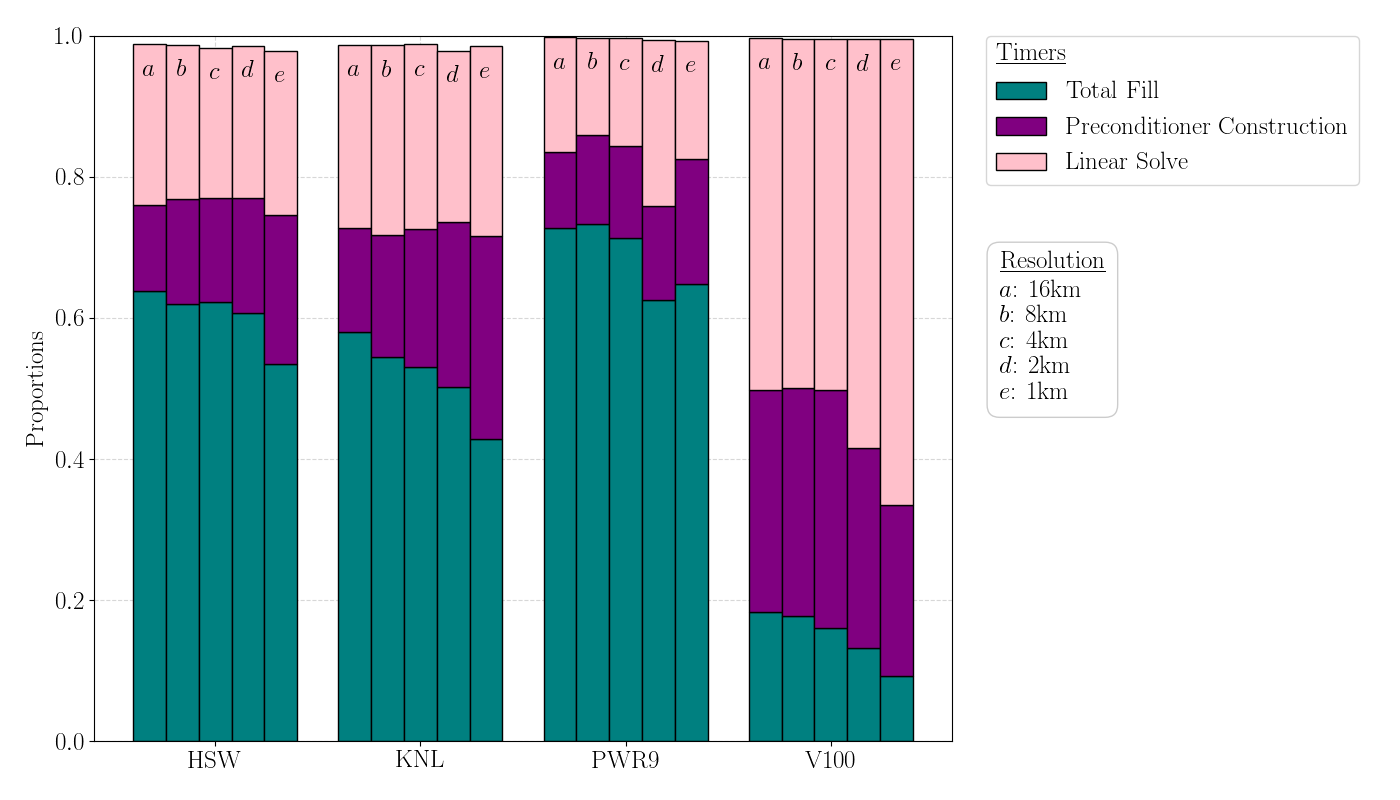}
\caption{A series of increasingly higher resolution Antarctic ice-sheet simulations are executed in a weak scalability study on four architectures (Cori: HSW, KNL; Summit: PWR9, V100). Timers are defined in Table~\ref{tab:res-timers}. This plot shows the ratio of each timer compared to \textbf{Total Solve}.}
\label{fig:res-weak-prop}
\end{figure}

Figure~\ref{fig:res-weak-prop} shows the proportions of total wall-clock for each architecture, resolution and timer.
On CPU platforms, \textbf{Total Fill} is the dominant contributor to \textbf{Total Solve} performance across all resolutions. At lower resolutions, \textbf{Preconditioner Construction} becomes a larger contributor. On GPU platforms, it's clear that \textbf{Linear Solve} is the largest contributor across all resolutions with \textbf{Total Fill} falling to less than 10\% at the lowest resolution.

\subsection{ALI Greenland ice-sheet 1-to-7 kilometer variable resolution performance test} \label{subsec:res-test}
\begin{figure}[htbp]
\centering
\begin{subfigure}{\textwidth}
\centering
\includegraphics[width=\textwidth]{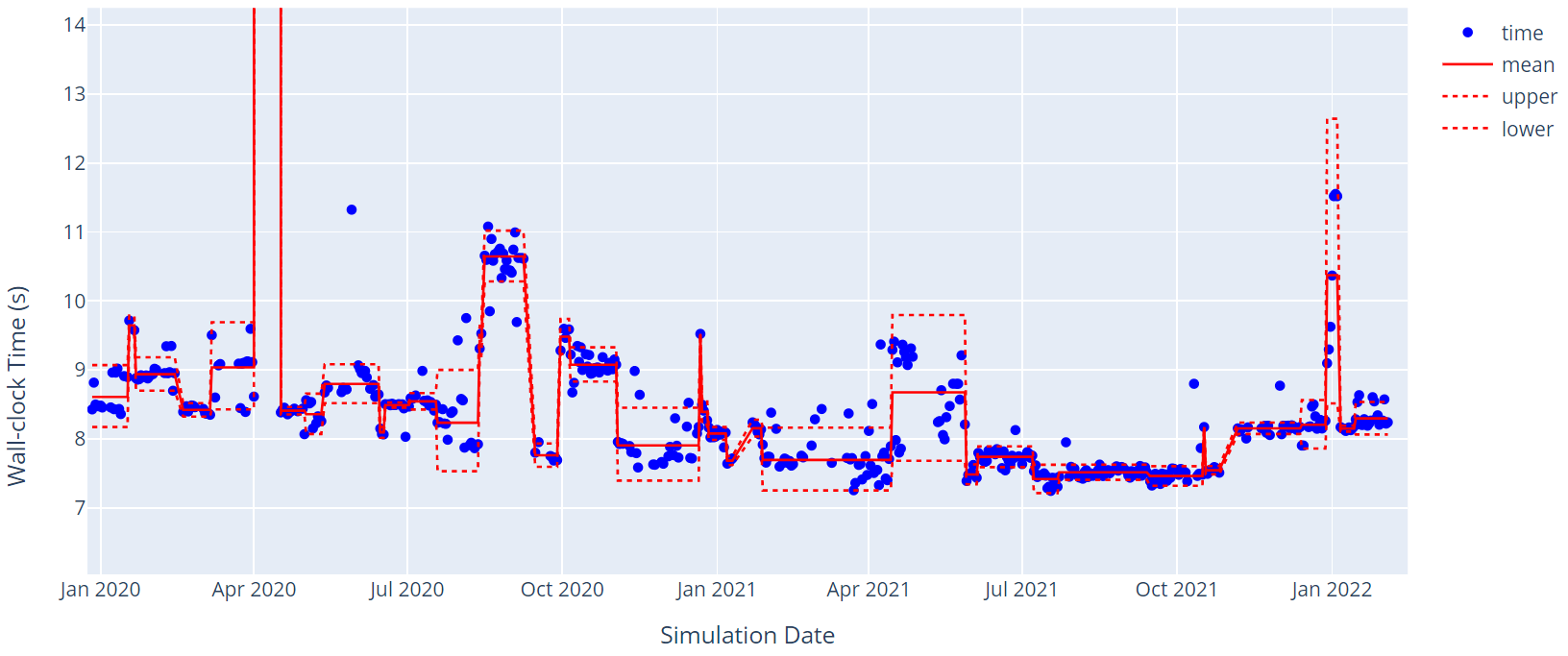}
\caption{\textbf{Total Time} on 8 Blake nodes (384 Skylake cores)}
\label{fig:res-test-blake-total}
\end{subfigure}
\vskip\baselineskip
\begin{subfigure}{\textwidth}
\centering
\includegraphics[width=\textwidth]{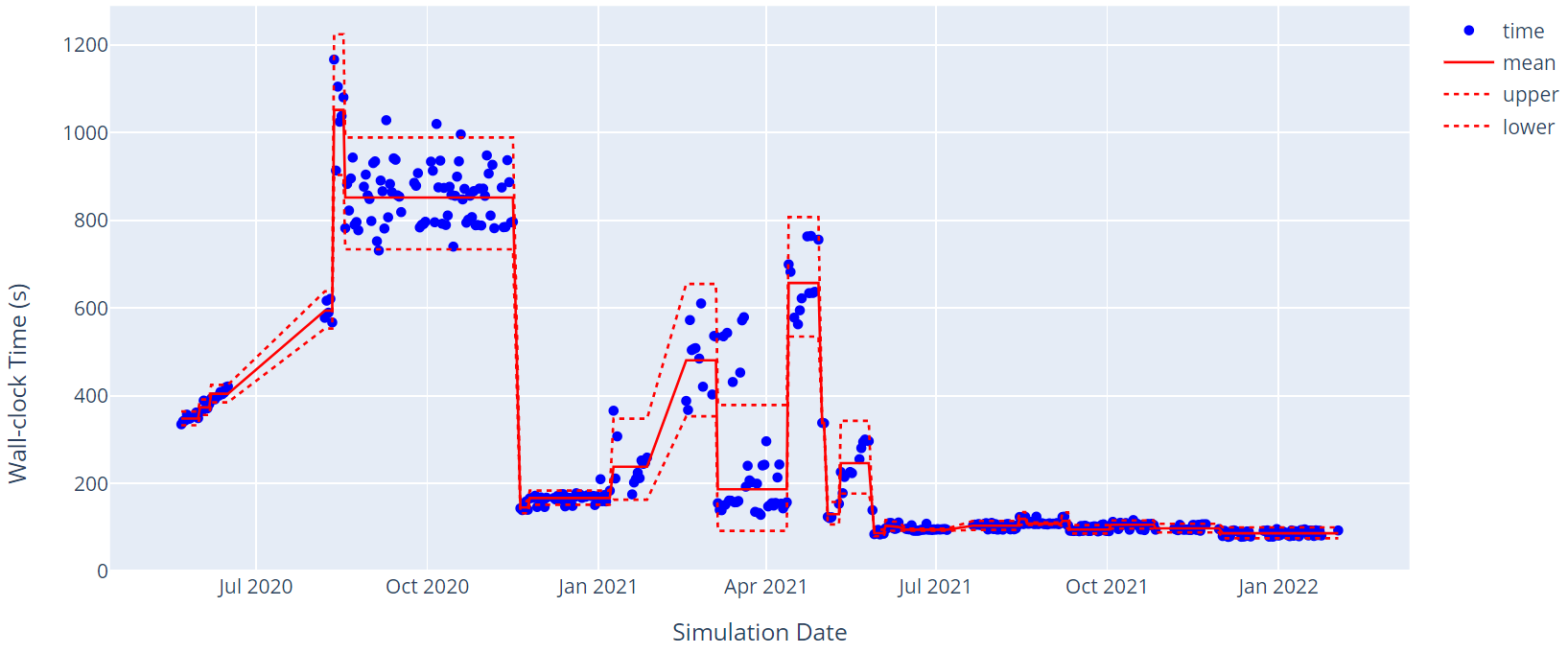}
\caption{\textbf{Total Time} on 2 Weaver nodes (8 V100 GPUs)}
\label{fig:res-test-weaver-total}
\end{subfigure}
\caption{The ALI Greenland ice-sheet 1-to-7 km variable resolution simulation is executed nightly on two platforms in order to detect regressions and improvements. Blue markers are recorded wall-clock time. Means are computed between changepoints and are indicated by solid red lines. The dotted red lines are $\pm$ two standard deviations.}
\label{fig:res-test-total}
\end{figure}

The last case focuses on solving the first-order velocity equations for a Greenland ice-sheet, 1-to-7 kilometer variable resolution mesh in a nightly performance testing framework for ALI. The numerical test is used to identify performance regressions and improvements within ALI. In this test, a two-dimensional, unstructured, Greenland ice-sheet mesh with 479,930 triangle elements is used. The test first extrudes the mesh by 10 layers using 3 tetrahedra per layer to create a mesh with 14,397,900 elements and 5,520,460 degrees of freedom. Then, equation~\eqref{eq:FOStokes} is solved using the methods described in Section~\ref{sec:impl}, and the mean value of the final solution is compared to previously tested values using a relative tolerance of \num{1.0e-5}.
The basal sliding coefficient is estimated using deterministic inversion from observed surface velocities \cite{perego2014optimal} and a realistic temperature field is provided. The test is currently running on two small clusters with different HPC architectures as shown in Table~\ref{tab:res-test-clusters}.
\begin{table}[htbp]
\caption{ALI nightly performance tests are executed nightly on the two small clusters given below.}
\label{tab:res-test-clusters}
\begin{tabular}{p{3.0cm}p{6.0cm}p{6.0cm}}
\hline\noalign{\smallskip}
Name & Blake & Weaver \\
\noalign{\smallskip}\hline\noalign{\smallskip}
CPU & Intel Xeon Platinum 8160 \newline Skylake & IBM POWER9 \\
Number of Cores & 24 & 20 \\
GPU & None & NVIDIA Tesla V100 \\
Node Arch. & 2 CPUs & 2 CPUs + 4 GPUs \\
Memory per Node & 188 GiB & 319 GiB + 15.7 GiB/GPU \\
CPU Compiler & Intel 18.1.163 & gcc 7.2.0 \\
GPU Compiler & None & nvcc 10.1.105 \\
MPI & openmpi 2.1.2 & openmpi 4.0.1 \\
Node Config. & 48 MPI & 4 MPI \\
\noalign{\smallskip}\hline\noalign{\smallskip}
\end{tabular}
\end{table}
The simulations executed on Blake and Weaver utilize 8 and 2 nodes, respectively. The historical \textbf{Total Time} for the two cases is shown in Figure~\ref{fig:res-test-total}.
The plots show variability associated with the code base and the system. Statistically significant changepoints are detected using the methods described in Section~\ref{subsec:impr-test} in order to identify regressions and improvements. The simulations in both time series utilize memoization so the CPU performance in Figure~\ref{fig:res-test-blake-total} has not changed much. In contrast, many of the other changes described in section~\ref{sec:impr} have been added over the course of the time series causing dramatic improvements to GPU performance as shown in Figure~\ref{fig:res-test-weaver-total}.

\begin{figure}[htbp]
\centering
\begin{subfigure}{0.48\textwidth}
\centering
\includegraphics[width=\textwidth]{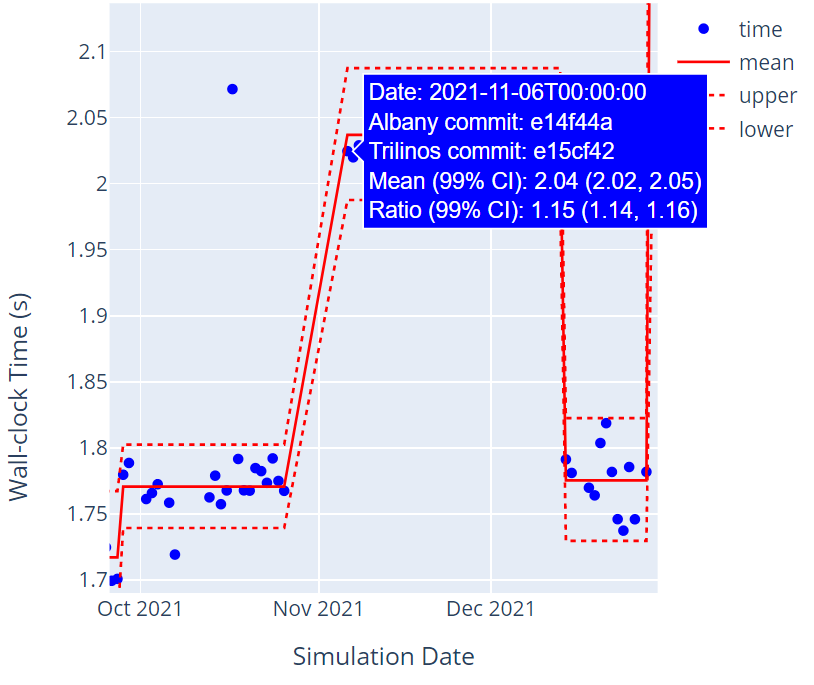}
\caption{\textbf{Total Fill} regression}
\label{fig:res-test-blake-fill-regr}
\end{subfigure}
\hfill
\begin{subfigure}{0.48\textwidth}
\centering
\includegraphics[width=\textwidth]{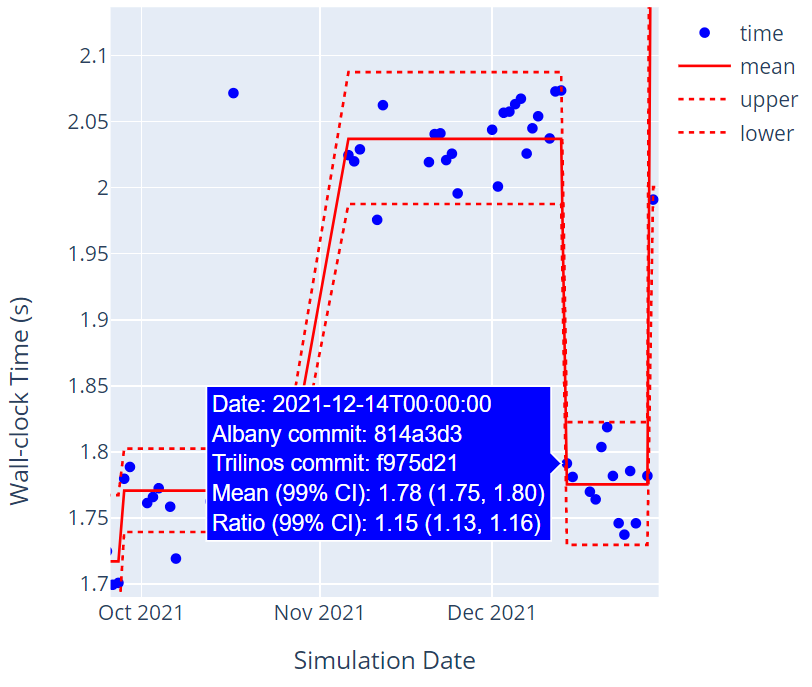}
\caption{\textbf{Total Fill} improvement}
\label{fig:res-test-blake-fill-impr}
\end{subfigure}
\caption{The ALI Greenland ice-sheet 1-to-7 km variable resolution simulation is executed nightly on Blake in order to detect regressions and improvements. Blue markers are recorded wall-clock time. Means are computed between changepoints and are indicated by solid red lines. The mean values are also given in the blue box along with a 99\% confidence interval (CI). CIs are reported as (LL, UL) where LL is the lower limit and UL is the upper limit. The dotted red lines are $\pm$ two standard deviations. Lastly, the blue boxes also show ratios (speedup, slowdown) between sets with a 99\% CI. This particular case highlights a performance regression and its later improvement from the fix.}
\label{fig:res-test-blake-fill}
\end{figure}

Figure~\ref{fig:res-test-total} also shows many performance regressions and improvements over the course of the time series. One recent example is the transition to Kokkos 3.5.0 which caused a regression to CPU performance. The regression along with the improvement from the fix is shown in Figure~\ref{fig:res-test-blake-fill}.
In this particular case, the usage of Kokkos \lstinline{atomic_add} was changed for the \lstinline{Serial} execution space. This caused a 15\% (99\% CI: 14\%, 16\%) slowdown to \textbf{Total Fill}. This was then fixed manually in Albany by avoiding the use of \lstinline{atomic_add} when running with a \lstinline{Serial} execution space. After the change, Figure~\ref{fig:res-test-blake-fill-impr} shows that the performance improved by 15\% (99\% CI: 13\%, 16\%), reverting the performance loss due to the original regression. Since then, the issue was reported and a fix has been introduced into Kokkos.

\begin{figure}[htbp]
\centering
\begin{subfigure}{\textwidth}
\centering
\includegraphics[width=\textwidth]{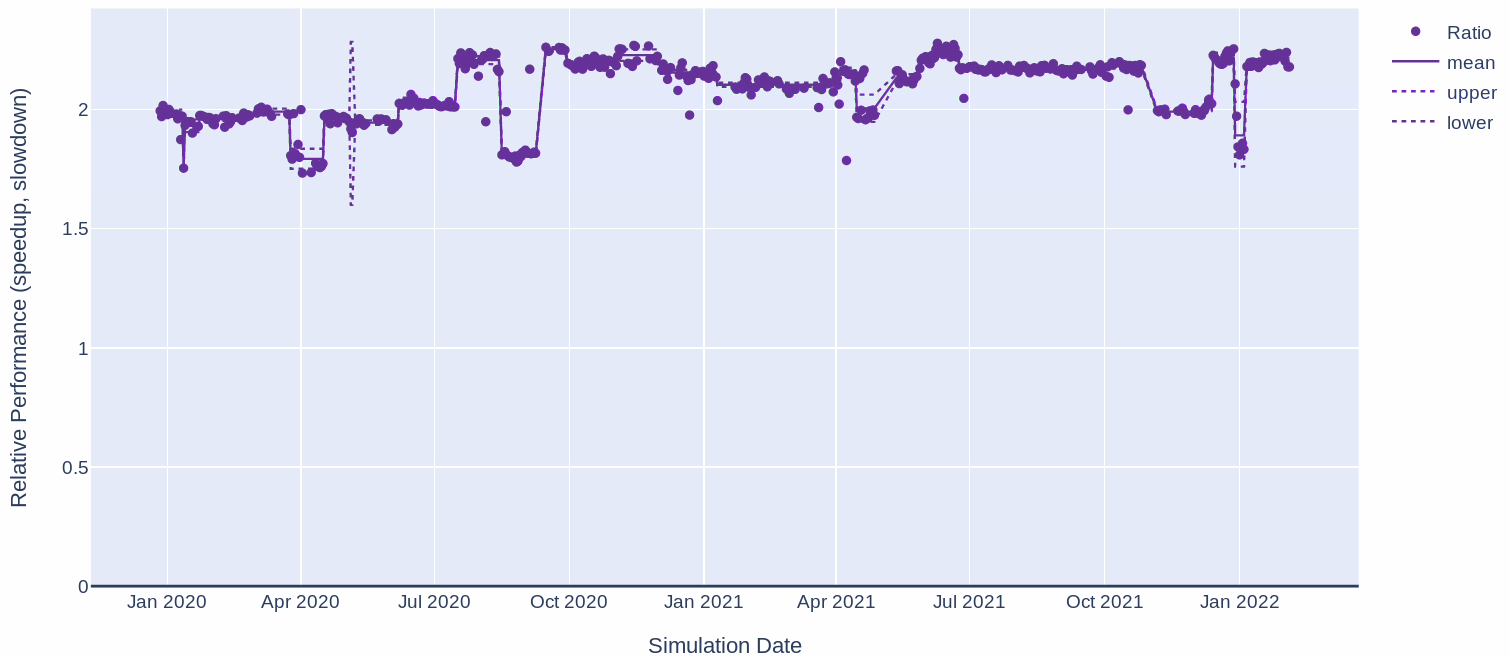}
\caption{\textbf{Total Fill} on 8 Blake nodes (384 Skylake cores)}
\label{fig:res-test-blake-fill-comp}
\end{subfigure}
\vskip\baselineskip
\begin{subfigure}{\textwidth}
\centering
\includegraphics[width=\textwidth]{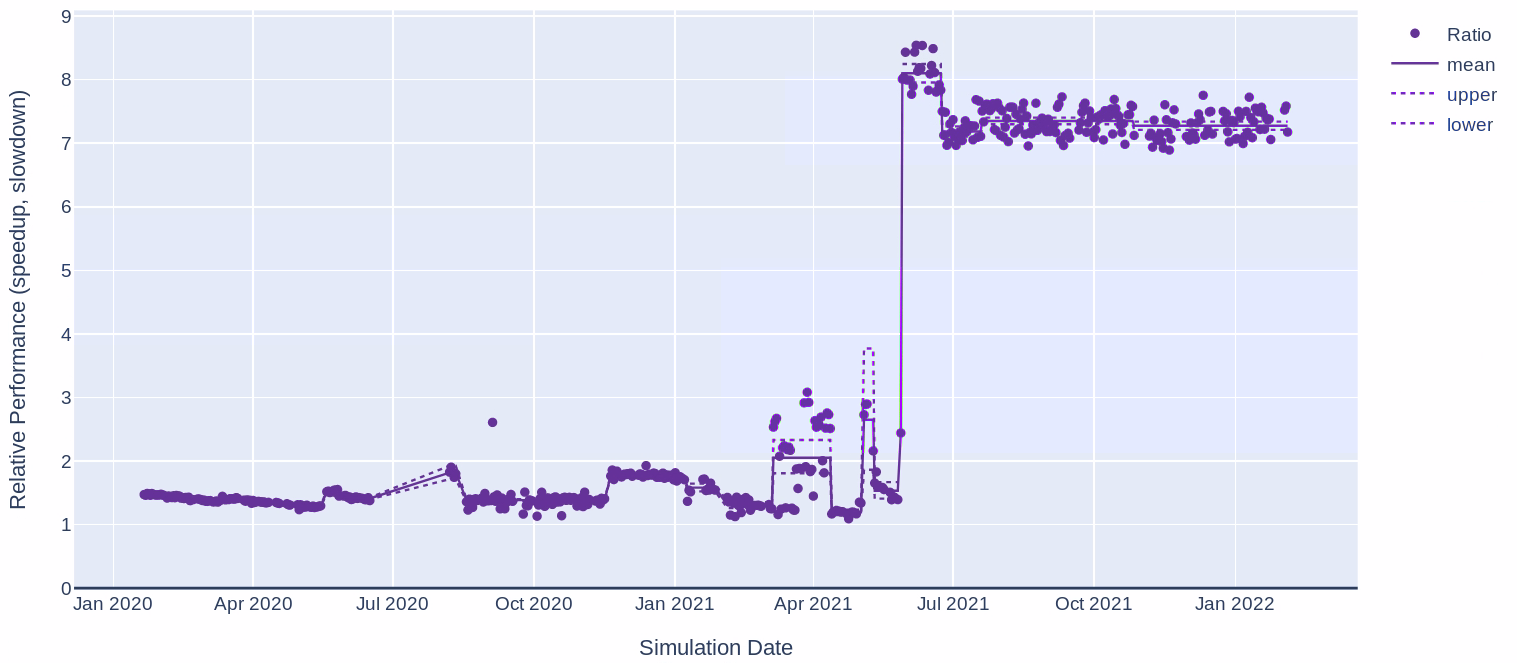}
\caption{\textbf{Total Fill} on 2 Weaver nodes (8 V100 GPUs)}
\label{fig:res-test-weaver-fill-comp}
\end{subfigure}
\caption{The ALI Greenland ice-sheet 1-to-7 km variable resolution finite element assembly tests with and without memoization are executed nightly on two platforms in order to detect regressions, improvements and analyze comparisons. Observations from the two cases are joined by date by taking the difference between the log of the timer data and plotting the relative performance (speedup, slowdown) with markers. Solid lines indicate means between changepoints and dotted lines represent a 99\% confidence interval for the mean.}
\label{fig:res-test-fill-comp}
\end{figure}

Algorithm performance comparisons can also be historically tracked within the performance testing framework. One example is the performance comparison of the finite element assembly with and without memoization. In this case, two performance tests which store the wall-clock time of 100 residual and Jacobian evaluations are tracked. Figure~\ref{fig:res-test-fill-comp} shows the observations between the two cases paired by simulation date on Blake and Weaver.
Both plots show that the test with memoization has performed faster than the test without memoization throughout the entire time series. On the CPU platform shown in Figure~\ref{fig:res-test-blake-fill-comp}, the relative performance has not changed much but on the GPU platform in Figure~\ref{fig:res-test-weaver-fill-comp} the relative performance has increased over time. The key difference in this case was the boundary condition improvements which significantly reduced the \textbf{Total Fill} time and caused evaluators without memoization to take up a larger portion of the \textbf{Total Fill} time.

\begin{figure}[htbp]
\centering
\begin{subfigure}{0.48\textwidth}
\centering
\fbox{\includegraphics[width=0.95\textwidth]{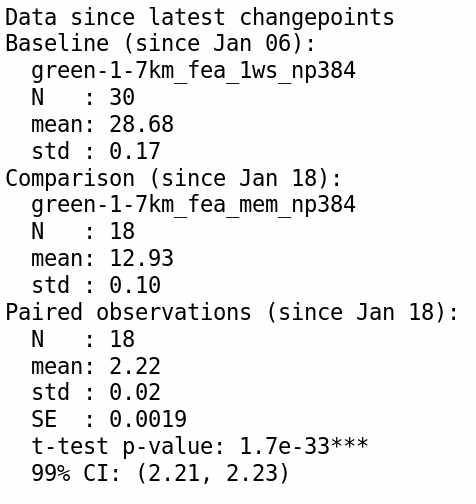}}
\caption{\textbf{Total Fill} on 384 Skylake cores}
\label{fig:res-test-blake-fill-comp-CI}
\end{subfigure}
\hfill
\begin{subfigure}{0.48\textwidth}
\centering
\fbox{\includegraphics[width=0.95\textwidth]{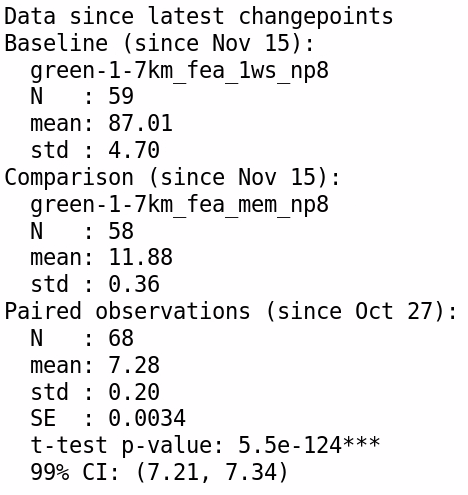}}
\caption{\textbf{Total Fill} on 8 V100 GPUs}
\label{fig:res-test-weaver-fill-comp-CI}
\end{subfigure}
\caption{The ALI Greenland ice-sheet 1-to-7 km variable resolution finite element assembly tests with and without memoization are executed nightly on two platforms in order to detect regressions, improvements and analyze comparisons. Observations from the two cases are joined by date by taking the difference between the log of the timer data and performing a paired $t$-test. The figures show sample sizes since the last changepoint for each test as well as mean wall-clock times (s) and standard deviation. The paired relative performance (speedup) is also given along with the standard error, p-value and a 99\% confidence interval (CI). CIs are reported as (LL, UL) where LL is the lower limit and UL is the upper limit.}
\label{fig:res-test-fill-comp-CI}
\end{figure}

The time series since the most recently detected changepoint can be used to determine whether the latest relative performance for memoization is statistically significant. A paired $t$-test is used to test the mean difference and the data is summarized as shown in Figure~\ref{fig:res-test-fill-comp-CI}.
The results show that the current estimated speedup from memoization for this case is 2.22 (99\% CI: 2.21, 2.23) on CPU platforms and 7.28 (99\% CI: 7.21, 7.34) on GPU platforms.

\section{Conclusions} \label{sec:conc}
In this paper, the performance portable features of MALI are introduced and analyzed on the two supercomputing clusters: NERSC Cori and OLCF Summit. First, the first-order velocity model and the mass continuity equation are introduced along with their implementations within Albany Land Ice and MPAS, respectively. This is used to further describe improvements that have been made to the finite element assembly process and linear solve within MALI. The new features focus on improving performance portability in MALI but are extensible to other applications targeting HPC systems.

Two numerical experiments are provided to analyze the expected performance on different HPC architectures. The first case utilized MALI to simulate an initial state calculation and single time step for a Greenland ice sheet 1-to-10 kilometer resolution mesh and compared baseline simulations without specific features with improved simulations with the features described in the paper. The results show that finite element assembly with memoization, MDSC-Kokkos and tuned smoothers are performant across all architectures with an expected speedup of 1.60 (99\% CI: 1.32, 1.93) on Cori-Haswell, 1.82 (99\% CI: 1.78, 1.86) on Cori-KNL, 1.26 on Summit-POWER9 and 1.30 (99\% CI: 1.29, 1.32) on Summit-V100. The study also highlights specific regions in need of improvement in the model. In particular, the need to improve the performance of the coupling between MPAS and Albany, the finite element assembly process on CPUs, and the preconditioner on GPUs.

The second numerical experiment utilized ALI to perform a steady-state simulation of the Antarctic ice sheet on a series of structured meshes in a weak scalability study. The results show that simulations on Summit-V100 perform the best with a 1.92 (99\% CI: 1.91, 1.92) speedup over Summit-POWER9 in the low resolution case and 1.24 (99\% CI: 1.21, 1.28) speedup over Summit-POWER9 in the high resolution case. The best results on Summit are shown during finite element assembly where the speedup over Summit-POWER9 is 8.65 (99\% CI: 8.22, 9.10). The results also show good weak scaling in finite element assembly for CPU/GPU but poor weak scaling in the preconditioner on CPU/GPU and in the linear solve on GPU architectures. Further analysis shows that the average number of linear iterations per nonlinear iteration increases dramatically as the resolution increases, highlighting the need for a more scalable preconditioner for this particular problem.

This paper also introduces a changepoint detection method for automated performance testing. A detailed description of the method is given along with examples of how the method can be used to detect performance regressions, improvements and differences in algorithm performance over time. In this case, an automated performance testing framework is used with ALI to simulate the Greenland ice sheet using a 1-to-7 kilometer variable resolution mesh. The results show the method being exercised on two years of data and an example of a successful detection of performance regression and improvement. The results also show an example of a nightly performance comparison where two tests are used to compare ALI with and without memoization. This case was used to show how the method can be used to detect regressions and improvements in algorithm performance over time as the utility of memoization has improved to up to 7.28 (99\% CI: 7.21, 7.34) speedup over simulations without memoization on GPU platforms over the course of two years.

\section*{Data availability}
The performance testing framework and data is available in \url{https://github.com/sandialabs/ali-perf-tests} and \url{https://github.com/sandialabs/ali-perf-data}. The results are accessible via a browser here: \url{https://sandialabs.github.io/ali-perf-data}. Performance data, pre-processing scripts and post-processing scripts are available in \url{https://github.com/sandialabs/ali-perf-data} under the directory paper-data/watkins2022performance.

\section*{Acknowledgments}
Support for this work was provided through the SciDAC projects FASTMath and ProSPect, funded by the U.S. Department of Energy (DOE) Office of Science, Advanced Scientific Computing Research and Biological and Environmental Research programs. This research used resources of the National Energy Research Scientific Computing Center (NERSC), a U.S. Department of Energy Office of Science User Facility operated under Contract No. DE-AC02-05CH11231. This research used resources of the Oak Ridge Leadership Computing Facility at the Oak Ridge National Laboratory, which is supported by the Office of Science of the U.S. Department of Energy under Contract No. DE-AC05-00OR22725.

The authors thank Trevor Hillebrand from Los Alamos National Laboratory for help with setting up the ice-sheet grids, datasets and for fruitful discussions. The authors also thank Luc Berger, Christian Glusa, Mark Hoemmen, Jonathan Hu, Brian Kelley, Jennifer Loe, Roger Pawlowski, Siva Rajamanickam, Chris Siefert, Raymond Tuminaro and Ichitaro Yamazaki from Sandia National Laboratories for their help with Trilinos components and Si Hammond for troubleshooting on Sandia HPC systems.

\section*{Disclaimer}
Sandia National Laboratories is a multimission laboratory managed and operated by National Technology and Engineering Solutions of Sandia, LLC, a wholly  owned subsidiary of Honeywell International, Inc., for the U.S. Department of Energy’s National Nuclear Security Administration under contract DE-NA-0003525.

This paper describes objective technical results and analysis. Any subjective views or opinions that might be expressed in the paper do not necessarily represent the views of the  U.S. Department of Energy or the United States Government.

\bibliographystyle{plain}
\bibliography{references}

\end{document}